\documentclass[11pt,twoside]{book} 

\usepackage{asp2008s}
\usepackage{times}
\usepackage{lscape}
\usepackage{epsf}

\usepackage{graphicx}
\usepackage{amssymb}
\usepackage{longtable}
\usepackage[figuresright]{rotating}
\usepackage{multirow}

\newcommand{\simless}{\mathbin{\lower 3pt\hbox {$\rlap{\raise 5pt\hbox{$\char'074$}}\mathchar"7218$}}}

\mainmatter

\newlength{\deftabcolsep}
\begin{document}


\title{Star Formation in the $\rho$ Ophiuchi Molecular Cloud}

\author{Bruce A. Wilking}
\affil{Department of Physics and Astronomy, University of Missouri-St.  Louis,
1 University Boulevard, St.  Louis, MO 63121, USA}

\author{Marc Gagn{\' e}}
\affil{Department of Geology and Astronomy, West Chester University, West Chester, PA 19383, USA}

\author{Lori E. Allen}
\affil{Harvard-Smithsonian Center for Astrophysics, 60 Garden Street, MS42, Cambridge, MA  02138, USA}

\begin{abstract}
A review of star formation in the Rho Ophiuchi molecular complex is presented, with
particular emphasis on studies of the main cloud, L1688, since 1991.  Recent photometric and parallax
measurements of stars in the Upper Scorpius subgroup of the Sco-Cen OB association suggest a
distance for the cloud between 120 and 140 parsecs.  Star formation is ongoing in the dense cores of
L1688 with a median age for young stellar objects of 0.3 Myr.  The surface population appears
to have a median age of 2-5 Myr and merges with low mass stars in the Upper Scorpius subgroup.
Making use of the most recent X-ray and infrared photometric surveys and spectroscopic surveys of
L1688, we compile a list of over 300 association members with counterparts in the 2MASS catalog.
Membership criteria, such as lithium absorption, X-ray emission, and infrared excess, cover
the full range of evolutionary states for young stellar objects.  Spectral energy distributions
are classified for many association members using infrared photometry obtained from the {\it Spitzer
Space Telescope}.

\end{abstract}

\section{Introduction}

The $\rho$ Ophiuchi cloud complex has been a target of intense studies in the past decade
due to some unique features.  First, it is one of the closest star-forming regions
with a distance estimated to be about 130 pc (see Section 2) and it is available
to observers in both the northern and southern hemispheres.  This proximity makes even
the low luminosity substellar objects in the region accessible to spectroscopy as
well as imaging.
Its large centrally condensed core, with gas column densities corresponding to A$_v$ =
50-100 mag, provides an effective screen against background stars and is an intermediate
example of star-formation relative to the low mass stars and isolated dense cores in
Taurus-Auriga clouds
and rich stellar clusters and massive cores in the Orion Molecular Cloud.

\begin{figure}
\begin{center}
\includegraphics[width=5in,draft=False]{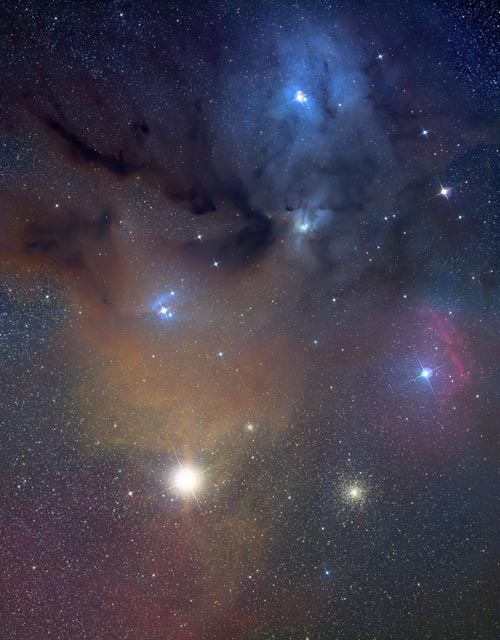}
\end{center}
\caption{An optical image of the Ophiuchus cloud complex and its environs taken by Robert Gendler,
Jim Misti, and Steve Mazlin.  The image is oriented with north pointing up and east to the left
and covers approximately 4.5 $\times$ 6 degrees surrounding the
dark cloud complex.  The bright threesome of stars in the northernmost blue reflection nebula
is the quadruple star system $\rho$ Oph from which the cloud
derives its name.  HD~147889 illuminates the blue reflection nebula immediately
west of the main dark cloud, L1688.  22~Sco is the pair of stars just east of center between L1688
and the wispier cloud, L1689.  The large red reflection nebula in the south is illuminated
by the M supergiant $\alpha$ Sco.}
\end{figure}

The basic structure of the cloud complex and its surroundings can be
seen in the color image shown in Fig. 1.  B stars in the Sco-Cen association,
as well as the M supergiant
$\alpha$ Sco, illuminate the dust bordering the dark dust clouds.
The B2V star HD~147889 lies
on the western edge of the centrally condensed core and the main cloud L1688.
Dark streamers trail to the northeast
(L1709) and the southeast (L1689).  The detailed extinction due to dust
derived from star counts has been presented by \citet{Cambresy1999} and \citet{Lombardi2008}
and the corresponding column density of molecular gas can be seen in the
$^{13}$CO integrated intensity maps
obtained for the {\it COMPLETE Survey of Star-Forming Regions} (see Fig. 2).

\begin{figure}
\begin{center}
\includegraphics[width=5.7in]{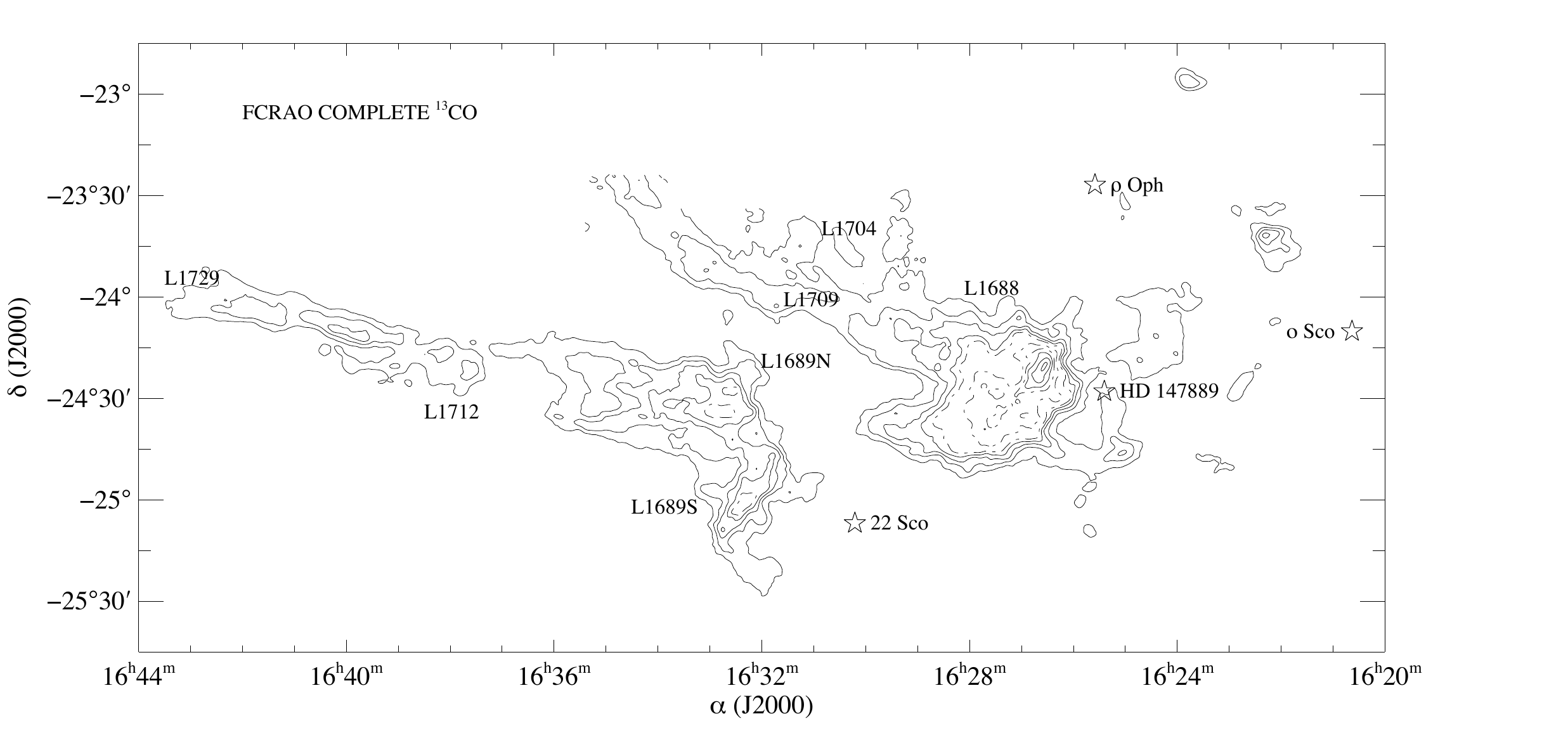}
\end{center}
\caption{A map of $^{13}$CO integrated intensity in the Ophiuchus cloud complex taken from the
COMPLETE survey \citep{Ridge2006}.
The major clouds are labeled
with their Lynds designations as are four B stars from the Upper Scorpius subgroup of the Sco-Cen
OB association.}
\end{figure}

Since the last review of this cloud prepared in 1991 \citep{Wilking1991}, there has been a wealth of
new instruments and data available on the Ophiuchus region.  At high energies,
the ROSAT, Chandra, and XMM-Newton X-ray telescopes have surveyed this region, displaying
a high sensitivity to young stellar objects (YSOs).  Large aperture ground-based
telescopes have opened the possibility for spectroscopic studies of very low mass and
substellar objects.  Ground-based near-infrared camera surveys by various groups and
2MASS have given us for the first time a complete sample of sources over a large area.
With the repair of the NICMOS camera on the Hubble Space Telescope, deep near-infrared
imaging from space has been possible.  The space-based infrared telescopes, the Infrared Space Observatory
and the Spitzer Space Telescope,
have surveyed the cloud, sensitive to warm dust surrounding YSOs. And finally array
receivers on millimeter-wave telescopes have made large scale surveys of dense gas and
dust possible.

In this chapter the main focus is a review of these new data sets and the basic results that have
advanced our knowledge of this star-forming region since 1991 through 2007.  The review
concentrates on the main cloud, L1688.  In Section 2, we review the relationship of L1688
to the Sco-Cen OB association and distance estimates to the $\rho$ Oph cloud.  Recent
molecular-line and millimeter continuum observations that have revealed the clumpy
structure of the cloud are discussed in Section 3.  In Section 4, we review
observations covering the entire electromagnetic spectrum that have enabled the
identification of many new YSOs in the cloud.  We present a table of over
300 YSOs associated with the cloud, along with their positions and infrared magnitudes
from the 2MASS survey.  In addition, we present a table of the most heavily
obscured YSOs, some too faint to be detected at near-infrared wavelengths, and
a table of candidate young brown dwarfs.  In Section 5, we briefly discuss
outflow activity in the cloud.  Finally, we review star formation in the
lower density streamers of the $\rho$ Oph cloud complex in Section 6.

\section{Relationship with Sco-Cen and Distance to L1688}

The idea that star formation in the L1688 cloud was triggered by compression from
the Sco-Cen OB association has a long history \citep{Vrba1977,LorenandWootten1986}.
Both cloud morphology and cloud chemistry has been cited as evidence.  \citet{deGeus1992}
has identified HI loops surrounding the Sco-Cen subgroups. He proposes that a slow shock
has been produced by an interaction between the expanding shell from the Upper Scorpius
subgroup and ambient gas from the L1688 cloud. The formation of this shell requires a
supernova event 1 to 1.5 Myr ago; the runaway star $\zeta$ Oph is proposed to be the
binary companion to the supernova \citep{Blauuw1961}.  This scenario is consistent with the
compression of the L1688 core and median age of 0.3 Myr for YSOs in the core
\citep{GreeneandMeyer1995,LuhmanandRieke1999}.
Further evidence for the triggering by Upper Scorpius comes from the alignment
of Class I protostars perpendicular to the proposed direction of shock propagation
\citep{MotteAndreandNeri1998}.

However, spectroscopic studies of YSOs from a broader region ($\sim$1 deg$^2$)
surrounding the L1688 cloud suggests a population of YSOs that predates the arrival
of this slow shock.  Spectroscopic studies by \citet{BouvierandAppenzeller1992},
\citet{Martinetal1998}, and \citet{Wilking2005} have characterized a population of
$\sim$100 YSOs that lie at the periphery of the L1688 cloud with a median age of 2 Myr.
Wilking et al. point out that an H-R diagram of these objects is indistinguishable
from that of low mass stars in the Upper Scorpius subgroup presented by
\citeauthor{Preibischetal2002}
(2002 and references therein).  Therefore, it is proposed that the formation of this
surface population of stars surrounding the L1688 core was contemporaneous with low
mass stars in the Upper Scorpius subgroup in a once larger L1688 cloud.  The age for
this earlier episode of star formation ranges from 2-5 Myr ago, depending on the correction
for unresolved binaries.  The trigger for this event, and the formation of the
Upper Scorpius subgroup, was presumably the passage of an expanding shell from the Upper
Centaurus-Lupus OB subgroup.  This sequence of events is consistent with the similar
velocities derived for young stars associated with $\rho$ Oph and Upper Sco \citep{Mamajek2007}.

The distance to the $\rho$ Ophiuchi molecular complex has undergone some revision in the past
decade from the canonical value of 160 pc \citep{Bertiau1958,Whittet1974}.  The
change has been due in large part to the Hipparcos Space Astrometry Mission and
accurate parallax values for stars in the Upper Scorpius subgroup.  Yet due to the
high visual extinction from the cloud, there are few bright stars known to be associated
with the cloud with well-determined parallaxes.
Based on ground-based photometry, \citet{deGeusetal1989} derived extinctions and absolute
magnitudes for stars in the Upper Scorpius subgroup.  From a plot of the extinction
vs. distance modulus, they defined the front edge of the cloud at 80 pc and the far
edge at 170 pc.  This yielded an estimate of 125 $\pm$ 25 pc for the cloud center.
\citet{deZeeuw1999} used Hipparcos proper motions
and parallaxes to define a group of 120 association members in the Upper Scorpius subgroup
over an area of 20\deg x 20\deg.  The mean distance derived for this group was 145 $\pm$ 2 pc.

\citet{deGeus1992} has noted that there are gaps in the HI emission that forms an expanding
shell originating from the Upper Scorpius subgroup and that these gaps correspond to
CO emission from the $\rho$ Oph cloud.  He proposed that the molecular cloud sits in
front of the HI shell.  This would imply a distance to the cloud closer than the mean
distance to Upper Scorpius.  Using the Hipparcos catalog trigonometric parallaxes,
\citet{KnudeandHog1998} compiled
color excesses and distances derived from the parallaxes for O-M stars within
6 degrees of the cloud.  They found an extinction jump at 120 pc and suggested an
upper limit of 150 pc to the cloud.
\citet{Mamajek2007} has estimated
a distance of 131$\pm$3 pc from a sample of seven association members within 5\deg\ of L1688.
\citet{Makarov2007} used astrometric proper motions for 58 probable association members over
the same region to derive a distance of 145 pc using the convergent point method.
Guided by an extinction map derived from 2MASS, \citet{Lombardi2008} used Hipparcos and Tycho parallaxes
combined with a maximum likelihood analysis to obtain a distance of 119 $\pm$ 6 pc.
These estimates include the star HD~147889 (136$\pm$26 pc) which appears to sit just behind the western edge
of the molecular cloud \citep{Liseau1999}.  Most recently, \citet{Loinard2008} have used VLBI
astrometry of two association members with strong radio emission
to derive a distance of 120.0 pc (+4.5 pc, $-$4.2 pc) for
the northern L1688 core, however, the same technique yields greater distances for sources
in the eastern core.  Clearly, the L1688 cloud exists over a range of distances, most likely
between 120 pc and 145 pc.
For this review, we adopt a distance of 130 pc
to L1688.

\section{Molecular Gas and Dust and Cloud Energetics}

The large scale structure of the $\rho$ Oph molecular complex had been established pre-1990
by extensive molecular line maps in $^{13}$CO with spatial resolutions of 2.4\arcmin\
\citep{Loren1989a,Loren1989b} or 2.7\arcmin\ (Mizuno et al. 1989, as quoted in
Nozawa et al. 1991).
More recently, high resolution (18\arcsec) observations of C$^{18}$O in the L1688 core have
been presented by \citet{Umemotoetal2002}.  A map of the C$^{18}$O column density obtained
at FCRAO with 46\arcsec\ resolution is shown in Fig. 3 (Li 2005, priv. communication).
These collective data
show a filamentary structure with gas column densities in the core trending along
a northwest-southeast axis and in the streamers along a perpendicular direction.

The large scale energetics of the cloud have been studied in greater detail
at far-infrared and submillimeter wavelengths with the Infrared Space Observatory
(ISO) and ground-based telescopes.  Far-infrared images by ISO confirm that the
heating of dust is dominated by B stars in the cloud (SR~3 and Oph Source 1) and
by the B2V star HD~147889 which lies at the western edge of the cloud
\citep{Abergeletal1996}.  Extended emission in the band centered on 6.7 $\mu$m appears
to be dominated by small grains and polycyclic aromatic hydrocarbons.
A bright filament of far-infrared emission corresponds to the western edge of the dense molecular
core and spectrophotometry with ISO of [CII] and [OI] lines show that this delineates
an ``edge-on" photodissociation region (PDR) associated with HD~147889 \citep{Liseau1999,Kulesa2005}.
Mapping of the [CI] and [CII] lines show that they primarily trace emission from the
lower density envelopes of molecular cores which are illuminated from the far side of the
cloud \citep{Kamegai2003,Kulesa2005}.  This emission mimics the column density distribution
of the molecular gas and is modeled with temperatures of 50-200 K
surrounding gas with temperatures of $\le$20K \citep{Kulesa2005}.

In the past decade, observations have focused more on the clumpy structure
of the densest gas and dust in the L1688
cloud using millimeter and submillimeter continuum, and molecular-line
observations.
Continuum mapping at $\lambda$= 3 mm, 1 mm, 850 $\mu$m, 800 $\mu$m,
450 $\mu$m, and 350 $\mu$m have revealed the clumpy structure in $\rho$ Oph core A
whose location is shown in Fig. 3 \citep{Ward-Thompsonetal1989,AndreWard-ThompsonandBarsony1993,
Wilson1999,DiFrancescoAndreandMyers2004}.  This structure includes three
cold cores, dubbed SM~1, SM~1N,
and SM~2, which appear to be prestellar cores each containing $\sim$0.5 M$_{\sun}$
as well as filaments and arcs which are likely related to outflows or PDRs.
Higher resolution interferometric mm continuum observations show that these
features contain sub-structures \citep[see][]{Kamazaki2001}.
More recently, wide field surveys have revealed the clumpy structure in L1688
at angular resolutions of 11\arcsec, 14\arcsec, 24\arcsec, and 31\arcsec,
respectively \citep{MotteAndreandNeri1998,Johnstone2000,
Smithetal2005,Young2006}.
These studies have identified about 55 starless cores in the range of 0.02-6.3 M$_{\sun}$
which comprise only a few percent of the total cloud mass.
The clump mass spectrum, dN/dM $\propto$ M$^{-\alpha}$, displays a slope of $\alpha\sim$1.5
for clumps with M$>$0.5 M$_{\sun}$ and a shallower slope of $\alpha$=0.5 for lower mass clumps.
This mass spectrum resembles the stellar initial mass function and it is suggested that
if each of the cores collapsed into individual stars, they could produce an
initial mass function.  The observed mass spectrum implies that turbulent fragmentation
at the prestellar stage plays a dominant role in its production \citep[see][]{Padoan2002}. This has been
confirmed through N$_2$H$^+$ observations which show that most of the clumps are
gravitationally bound and that the small velocity dispersion of the clumps implies insufficient
time for clump interactions prior to star formation \citep{Andre2007}.

\begin{figure}
\includegraphics[height=4.7in]{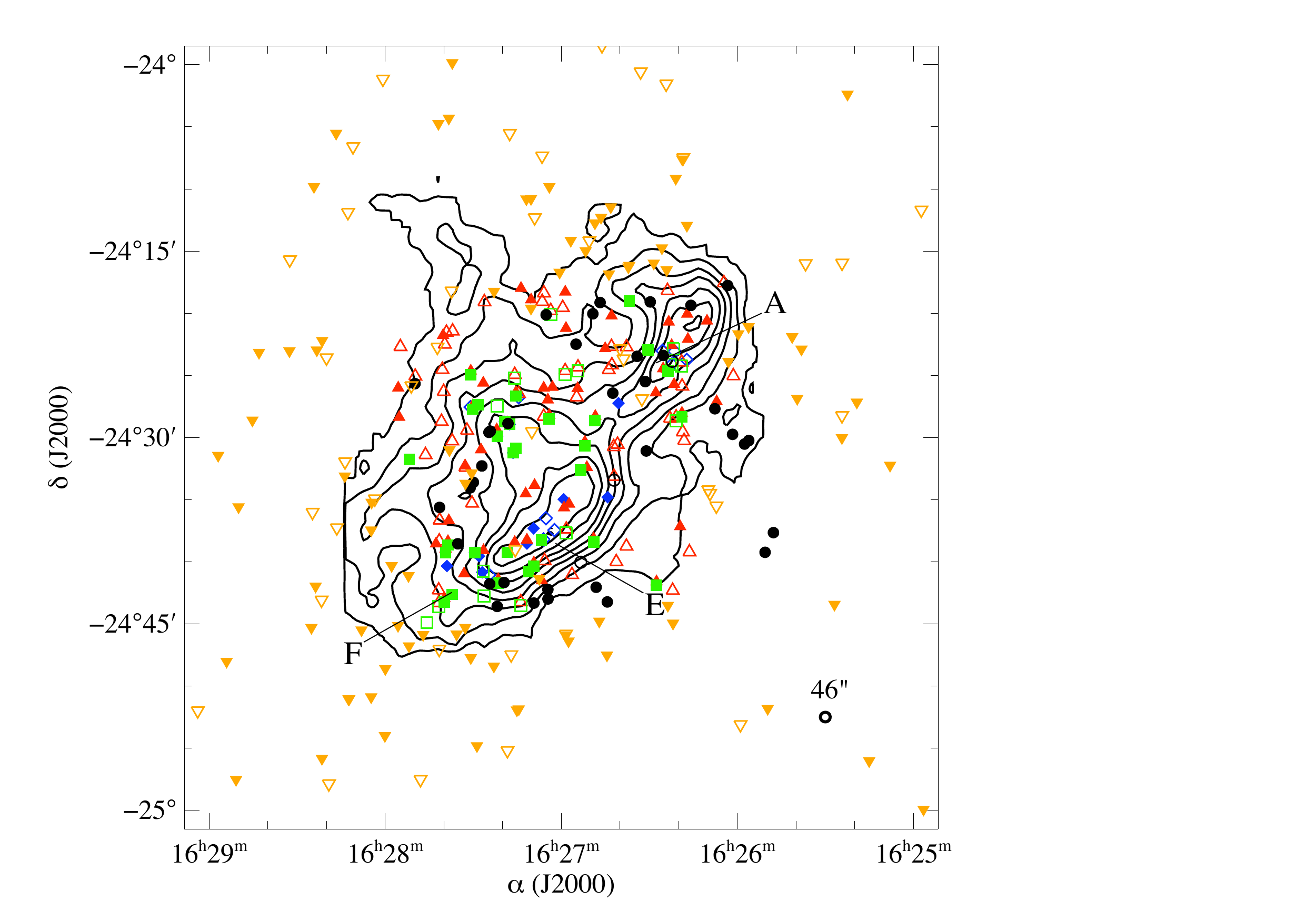}
\caption{The distribution of association members in L1688 also appearing in the
2MASS catalog relative to contours of C$^{18}$O integrated
intensity. The contours of integrated intensity start at 2 K km
sec$^{-1}$ and increase in steps of 1 K km sec$^{-1}$.  The general
locations of cold, dense cores A, E, and F (as traced by DCO$^+$
emission) are labeled and do not exactly correspond to the high column
density C$^{18}$O gas \citep[see
Fig. 1,][]{LorenWoottenandWilking1990}.  The source symbols indicate
the SED class for each association member as determined by their IRAC
colors: blue diamonds represent Class I objects, green squares
flat-spectrum sources, red triangles Class II objects, and black
circles Class III sources.  Inverted amber triangles have unclassified
SEDs.  Filled (open) symbols indicate that X-rays have been detected
(not detected) from the source.}
\end{figure}

\section{Young Stellar and Substellar Objects}

A list of association members in the L1688 cloud known up to January 2008
is presented in Table 1
which is available in Excel or IDL format.\footnote{Table 1 is available at either ftp://astro.wcupa.edu/pub/mgagne/roph or\\
 http://www.umsl.edu/$\sim$wilkingb/handbook/handbook.html
or by request to the authors.}
The area considered is a 0.94\deg $\times$ 1.2\deg\ box centered at
RA(2000) = 16$^h$ 27$^m$ 38\fs5, DEC(2000) = $-$24\deg 29\arcmin 34\arcsec.
All of the objects listed in Table 1 appear in the 2MASS catalog with,
at minimum, detections in the K$_s$ band.  In addition to their positions and
J, H, and K$_s$ magnitudes from the 2MASS catalog (col. 1-13), we list
source names from selected surveys (see Appendix A).
Association membership was based on various criteria which are listed in the second to last
column; the more criteria that a source satisfies, the more secure its
identification as a member.  X-ray emission (indicated by ``x" in Table 1)
is observed from young stellar objects in all evolutionary states (see Section 4.1).
While it is known to be observed from some main-sequence objects such
as dMe stars, the density of such objects over this region is
estimated to be low \citep{Gagne2004}.
YSOs observed spectroscopically (Section 4.3) could be identified by the presence of
strong H$\alpha$ emission, EW $>$ 10 \AA\ (``ha"), lithium absorption (``li"),
or location above the main sequence in the Hertzsprung-Russell diagram
\citep[``ext"; see][]{Wilking2005}.   Excess infrared emission is indicative
of a circumstellar disk and association membership (see Section 4.2).  Infrared excesses have
been established by combining near-infrared photometry with mid-infrared
photometry from ISO \citep[``IRX",][]{Bontempsetal2001,Comeronetal1998}
or Spitzer (B2ex, B4ex, or Mex).  Objects displaying an infrared excess in an (H$-$K$_s$) vs
(K$_s$$-$4.5 $\mu$m) diagram by being displaced by at least 2$\sigma$ from
the reddening line of an M9~V star are indicated by ``B2ex"
(IRAC Band 2 excess).  Objects whose IRAC colors (3.6$-$4.5 $\mu$m vs.
5.8$-$8.0 $\mu$m) fall in the regime of Class I or Class II sources
\citep{Allenetal2004} are indicated by ``B4ex" (IRAC Band 4 excess).  Sources with
spectral indices from 3.6 $\mu$m to 24 $\mu$m greater than $-$1.6 are noted by
``Mex" (MIPS excess).  Near-infrared variability is not listed as one of the criteria for association
but a recent study suggests additional association members may be identified through this
technique \citep{Alves2008}. Also noted in Table 1 is the classification of the spectral energy
distribution (SED) from either the placement in the
IRAC color-color diagram (1, F, 2) or the
3.6 $\mu$m to 24 $\mu$m spectral index (I, F, II).
The remaining objects have mainly Class III SEDs.

The distribution of association members listed in Table 1 is shown in Fig. 3
relative to contours of C$^{18}$O column density (Li 2005, priv. communication).

\subsection{X-ray Surveys}

The L1688 cloud was among the first regions imaged in X-rays by the
{\it Einstein} X-ray Observatory. \citet{Montmerle1983} discovered 70
highly variable X-ray sources in a $2\deg \times 2\deg$ field centered on
$\rho$~Oph core A.  In the X-ray surveys since with the {\it ROSAT}, {\it ASCA},
{\it XMM} and {\it Chandra} satellites, hundreds of X-ray sources have been
associated with Class I-III YSOs. Although YSOs emit less than 1\% of their
bolometric luminosity in the 0.1-10 keV soft X-ray band, X-ray surveys are a
good way to identify cloud members because the X-ray to bolometric luminosity
ratio of YSOs is much higher than most foreground and background field stars.
The absorption cross sections of H~I, He~I and He~II decrease rapidly with
increasing energy so that X-rays in the 2-10~keV band can often be detected
through $A_v=75$~mag of extinction.

\citet{Montmerle1983} found that the YSOs in $\rho$ Oph were continously flaring, producing
an X-ray ``Christmas Tree".
The X-ray spectra of YSOs show optically-thin thermal emission (a bremsstrahlung
continuum with bright emission of highly ionized metals).  Most YSOs in L1688 have
very hard, time variable X-ray spectra ($kT > 2$~keV) which suggest magnetic
heating of magnetically confined plasma. The extent and geometry of these magnetic
structures and the mechanisms that trigger flares and magnetic heating are still
under discussion.  Modeling of X-ray flares on Class I-III YSOs in $\rho$~Oph
\citep{Imanishi2003} suggest large magnetic loops ($\approx 1 R_\star$) and moderate
magnetic field strengths of 200-500~G.  Variability in fluorescent Fe 6.4 keV emission from the Class I
YSO Elias 29 has been interpreted as a magnetically confined accretion loop
extending from the star to the circumstellar disk \citep{Favata2005,Giardino2007}.
The presence of large magnetic structures have also been
suggested from VLBI observations of Oph Source 1, DoAr~21 (GSS~23), and
several lower mass Class III YSOs
\citep{Andreetal1991,Phillips1991,Andreetal1992}.

In this chapter we consider cloud members in a one square degree region
including the dense cores of L1688.  We note that many of the cloud members in this region
have been detected with {\it Einstein}, {\it ROSAT} \citep[see][]{Casanova1995,Martinetal1998,
Grosso2000} and {\it ASCA} \citep[see][]{Kamata1997}.  However, we focus on
the more sensitive, higher spatial resolution {\it Chandra} and {\it XMM} surveys
of cores A, E, and F by \citet{Imanishi2001}, \citet{Gagne2004}, and \citet{Ozawa2005}.  We
also examine surveys of
the outer regions for sources detected with {\it Chandra} (N. Grosso 2005,
private communication) or in the {\it XMM} DROXO survey \citep{Sciortino2006}.

Of the 316 candidate YSOs listed in Table 1, 201 have been detected in X-rays.
In cores A, E and F, where deep Chandra observations have detected 90\% of
candidate YSOs, the limiting X-ray luminosity is $\log L_{\rm X}\approx 28.3$.
The resulting X-ray luminosity function (XLF) for Class II and Class III sources
using the corrected distance of 130 pc is very similar to the XLF in the Orion
Nebula Cluster \citep{Feigelson2005}.
Assuming the Orion sample is complete and that the XLF is universal
\citep{Feigelson2005}, this result suggests that our census of Class II and Class III
objects in L1688 cores A, E, and F is essentially complete. In the outer regions
of the cloud, the X-ray detection fraction is lower, though the XLF accounts for
non-detections by using the Kaplan-Meier estimator.  The XLF in these outer regions
is statistically similar to the XLF in cores A, E and F, suggesting that the
candidate list (including X-ray non-detections) is relatively complete.

\subsection{Near-Infrared Surveys and Multiplicity}

Prior to 1992, most near-infrared surveys were conducted by scanning a single detector across
the dark cloud and were limited in sensitivity and spatial resolution.
Nevertheless, pioneering surveys by \cite{GrasdalenStromandStrom1973,Vrbaetal1975,Elias1978} and
\citet{WilkingandLada1983} revealed large numbers bright infrared sources in L1688
and multi-wavelength infrared studies established infrared excesses for many
of these sources \citep{Elias1978,LadaandWilking1984,WilkingLadaandYoung1989}.
Infrared arrays have revolutionized our view of star formation in nearby clouds and
L1688 is no exception.  In 1992, wide-field imaging surveys at J, H, and K began
to be published and gave us our first clear picture of the distribution of YSOs
across the cloud.  Some of the larger surveys are summarized below.  All of these were
precursors to the all-sky 2MASS survey which had a sensitivity at K of 14 mag
\citep{Cutri2003}.

\setcounter{table}{1}
\begin{table}[!ht]
\caption{Recent Targeted Near-Infrared Surveys in Ophiuchus}
\smallskip
\begin{center}
{\small
\begin{tabular}{lclr}
\tableline
\noalign{\smallskip}
JHK  Survey &          Instrument  &     Sensitivity  &         Coverage \\
\noalign{\smallskip}
\tableline
\noalign{\smallskip}
\citet{GreeneandYoung1992}        & UA NICMOS2  &  K$<$13 mag   &   650 arcmin$^2$ \\
\citet{Comeronetal1993}           & UA NICMOS2  &  K$<$15.5 mag &   200 arcmin$^2$ \\
\citet{StromKepnerStrom1995} & SQIID       &  K$<14.2$ mag &   1225 arcmin$^2$ \\
\citet{Barsonyetal1997}           & SQIID       &  K$<$14 mag   &   3660 arcmin$^2$ \\
\citet{Allen2002}             & HST NICMOS3 &  H$<$21.5 mag &   72 arcmin$^2$ \\
\noalign{\smallskip}
\tableline
\end{tabular}
}
\end{center}
\end{table}

One task of the wide-field surveys was to distinguish between association members and
background stars from JHK photometry alone.  Contamination of background stars could
be minimized if the surveys focused on the regions with the highest gas column
densities \citep[see][]{Comeronetal1993,Allen2002}.  Using a color-color diagram,
the near-infrared photometry could be plotted to look for infrared excesses at K due to
warm dust from the inner regions of a circumstellar disk and hence not only identify a
YSO but infer its evolutionary state.  A drawback of this technique is that it only detects
YSOs with inner disks and selects against weak-emission T Tauri stars (WTTS) and very low
mass objects.  High resolution images have identifed
YSOs by resolving dust structures around objects \citep{Allen2002}.  The more extensive
photometric
studies have been used to derive a luminosity function and infer a mass function for L1688
under the assumption of a common age and using a set of evolutionary models
\citep{Comeronetal1993,StromKepnerStrom1995}.  They have found that the mass function is
relatively flat below 1 M$_{\sun}$, a result confirmed through spectroscopic studies.

In addition to wide-field surveys, infrared and Gunn z surveys of individual
sources have been
performed to evaluate the binary frequency in L1688.  Techniques used include
direct imaging \citep{ReipurthandZinnecker1993,Simonetal1995,Brandner1996,
ResslerandBarsony2001,Haischetal2002,Duchene2004,
Haischetal2004}, speckle imaging \citep{Ghez1993,BarsonyKoreskoandMatthews2003,
Ratzka2005}, lunar occultations
\citep{Simonetal1987,Richichietal1994,Simonetal1995}, and adaptive optics
\citep{Ratzka2005,Duchene2007}.  YSOs which have companions within 10\arcsec\ are noted in the
last column of Table 1 provided the companions do not appear as separate entries.
The first studies focused on samples of
WTTSs and classical T Tauri stars (CTTSs) and
suggested an overabundance of multiple systems in the Ophiuchus complex compared
to the main-sequence.  More recent studies with larger samples find the degree of
multiplicity only marginally higher than that of main-sequence stars (Ratzka et al.
2005, see also Simon et al. 1995 and Haisch et al. 2002).  There are suggestions
that the degree of multiplicity drops with age when comparing Class I protostars and
flat-spectrum sources to main sequence stars \citep{Duchene2004} and CTTSs to WTTSs
\citep{Ratzka2005}.

\subsection{Spectroscopic Studies of YSOs and Brown Dwarfs}

In the previous review \citep{Wilking1991}, H$\alpha$ objective prism surveys were the main source of
spectroscopic information used to identify YSOs in the Ophiuchus region
\citep[e.g.,][]{DolidzeandArakelyan1959,WilkingSchwartzandBlackwell1987}.
Optical spectroscopy was available for a handful of CTTSs,
mainly the brightest sources with spectral types of K through early M \citep{StruveandRudkjobing1949,
CohenandKuhi1979,Rydgren1980}.  Comparisons of L1688 with
other star-forming regions showed a relative deficiency of mid-to-late M CTTSs in
L1688 \citep[see Fig. 24 in][]{Hillenbrand1997}; this was clearly a selection effect due
to the high visual extinction in the cloud.  In the past decade, it has become possible
to investigate large numbers of young stars in the Ophiuchus region with
high resolution and/or multi-object spectroscopy at both visible and near-infrared
wavelengths.  These studies can identify young stars through the presence of
hydrogen emission lines or CO rovibrational lines, or absorption by lithium.  Moreover, absorption lines
from species such as CaH, Na I, and Ca II can serve as surface gravity indicators.
Spectral classifications allow one to estimate effective temperatures and, when compared
with estimates of source luminosities and theoretical tracks and isochrones,
the masses and ages of YSOs.  Surface gravities derived from the spectra are also useful
in determining stellar ages.

\subsubsection{Optical Spectroscopy}

Currently, over 120 association members have been assigned spectral types using optical
spectroscopy (see Table 1).  However while optical spectra provide the most
reliable indicators of spectral type, they can only be obtained for low mass objects
with little or no visual extinction.
\citet{BouvierandAppenzeller1992} obtained spectral
types for 30 X-ray sources in and around the L1688 cloud and found most were
weak-emission T Tauri stars with ages ranging from 1 to 10 Myr.  \citet{Martinetal1998}
acquired optical spectral types for 59 X-ray sources observed over the entire Ophiuchus
complex.  They found the majority of objects classified were pre-main sequence (PMS)
M stars, with evidence for older PMS stars outside of the L1688 cloud.  Most recently,
\citet{Wilking2005} used a multi-object spectrograph to obtain spectra for 139 objects
within a 1.3 deg$^2$ area surrounding L1688.  A total of 88 association members were
identified with a large percentage possessing M spectral types.  The median age for this
sample is identical to that of low mass stars in the Upper Scorpius subgroup and
significantly older than those in the L1688 core.

\subsubsection{Near-Infrared Spectroscopy}

Because of the high visual extinction of objects in the L1688 core, significant
progress in the spectral classifications of association members has been made
at infrared wavelengths.  To date, over 90 association members have infrared spectral
classifications with only 30 of these having optical spectral classifications (see Table 1).
\citet{CasaliandMatthews1992} presented
low resolution K band spectra for 10 low-luminosity YSOs and showed
that those with Class II SEDs displayed $^{12}$CO absorption
bands.  \citet{GreeneandLada1996} presented moderate resolution 1.1-2.4 $\mu$m spectra
for a flux-limited sample (K$<$10.5 mag) of 53 YSOs in the L1688 cloud and showed
there was a relationship between the infrared SED
and the measured veiling
in the K band.  \citet{GreeneandMeyer1995} presented spectral types for 34 of these YSOs,
mainly those with Class II SEDs.   \citet{LuhmanandRieke1999}
conducted a deeper moderate-resolution spectroscopic survey (K$<$12 mag) in the K band
for $\sim$100 objects in L1688.  Using the models of \citet{DAntonaandMazzitelli1997},
a median age for YSOs in the L1688 core derived by these studies is 0.3 Myr.
\citeauthor{LuhmanandRieke1999} used these data
to construct an Initial Mass Function (IMF) for the L1688 core from 36 sources plus
a completeness correction for the lowest mass objects.  They found that for M$>$
0.4 M$_{\sun}$, the IMF was consistent with that of \citet{MillerandScalo1979}.
The IMF appears flat between 0.02 and 0.4 M$_{\sun}$ which agrees with previous
IR photometric and spectroscopic modeling \citep{Comeronetal1993,StromKepnerStrom1995,
Williamsetal1995} and with other young clusters such as IC 348 and
the Pleiades.

Using the SED class defined by ISO observations, \citet{Natta2006} obtained J and K band
spectra for 104 Class II objects with the goal of using the hydrogen emission
lines to estimate the accretion rate.  They have detected emission in $\sim$50\% of
the sample and find that accretion rate is strongly correlated with the mass
of the central object, although there is a large spread in accretion rates observed
for a given mass. There is no significant difference in this relationship
when objects in Ophiuchus are compared with those in Taurus for M$>$ 0.1 M$_{\sun}$.

Due to veiling, infrared echelle spectroscopy is required to resolve photospheric
absorption lines, particularly in the most deeply embedded YSOs.  In addition to enabling spectral
classification, observations of flat-spectrum and Class I protostars have confirmed
the higher degree of veiling (and presumably accretion rate) compared to CTTSs in L1688
\citep{GreeneandLada1997,GreeneandLada2000,GreeneandLada2002,
Doppmann2003,Doppmannetal2005}.  Stellar properties have been
determined for 16 highly veiled flat-spectrum or Class I protostars in the L1688 core
\citep{Doppmannetal2005}.  High resolution spectra have also
shown that the more deeply embedded sources rotate faster (and have higher
angular momenta) compared to CTTSs \citep{Coveyetal2005}.

\subsubsection{Spectroscopy of Brown Dwarf Candidates}

The proximity of the Ophiuchus complex and the enhanced luminosities of pre-main sequence
objects has made L1688 an ideal laboratory for the study of young brown dwarfs.
Given the median age of YSOs in the L1688 core, models suggest that objects with spectral
types of M6 or later are likely to be brown dwarfs.  Candidates were initially identified
through deep infrared imaging of the core \citep{RiekeandRieke1990,Comeronetal1993}.
As shown in Table 3,
confirmation of 20 objects as likely brown dwarfs has been made through low to moderate
resolution spectroscopy.  Three young brown dwarf candidates with spectral types of
M8-M8.5 have been identified with optical spectroscopy
\citep[CRBR~46, GY~3, and GY~264;][]{LuhmanLiebertandRieke1997,Wilking2005}
as well as three near the
hydrogen-burning limit
\citep[GY~5, GY~37, and GY~204;][]{WilkingGreeneandMeyer1999,Wilking2005}.
The remaining candidates have been observed using infrared
spectroscopy
\citep{Williamsetal1995,WilkingGreeneandMeyer1999,Cushing2000,Nattaetal2002}.
Spectral classifications
have been accomplished through comparisons of the depths of the water vapor and
other absorption bands with those of dwarf standard stars and with low surface gravity
synthetic spectra.  A significant result of these studies is that many of the brown
dwarf candidates in L1688 have near-to-mid infrared excesses indicative of circumstellar
disks \citep{Comeronetal1998,Testietal2002,Nattaetal2002}.
Analysis of hydrogen emission lines has
shown that very low mass objects and brown dwarfs in L1688 have lower accretion rates
than their more massive counterparts, but that brown dwarf candidates in L1688 exhibit
accretion rates that are generally an order of magnitude higher than objects of
comparable mass in other star-forming regions \citep{Nattaetal2004,Gatti2006}.

\begin{table}[!ht]
\caption{Brown Dwarf Candidates in L1688}
\smallskip
\begin{center}
{\small
\begin{tabular}{lcccc}
\noalign{\medskip}
\tableline
\noalign{\smallskip}
Name & Opt. Sp. Type & Ref. & IR Sp. Type & Ref.\tablenotemark{a} \\
\noalign{\smallskip}
\tableline
\noalign{\smallskip}
CRBR~14  &      &     & M7.5,M5.5,M7 & WGM99,LR99,N02 \\
CRBR~15  &      &     & M5           & WGM99,LR99 \\
GY~5     &M5.5  &  W05& M7,M6        & WGM99,N02 \\
GY~3     &M8    & W05 & M7.5         & N02 \\
GY~10    &      &     & M8.5,M6.5    & WGM99,LR99 \\
GY~11    &      &     & M6.5,M8.5    & WGM99,N02 \\
CRBR~23\tablenotemark{b}  &      &     & L1           & WMG07 \\
CRBR~31  &      &     & M6.7,M7      & CTK00,WMG07 \\
GY~31    &      &     & M5.5         & WGM99 \\
GY~37    & M5   & W05 & M6           & WGM99 \\
GY~59    & M3.75& W05 & M6,M5        & WGM99,LR99 \\
GY~64    &      &     & M8           & WGM99 \\
GY~84    &      &     & M6,M1.5      & WGM99,LR99 \\
GY~141   & M8.5 &LLR97& M8           & CTK00 \\
GY~201\tablenotemark{b}   & M9   & H00 & M5           & WMG07 \\
GY~202   &      &     & M7,M6.5      & WGM99,LR99 \\
GY~204   & M5.5 & W05 & M6           & N02 \\
GY~258   &      &     & M7           & WMG07  \\
oph-160  &      &     & M6           & N02 \\
GY~264   & M8   & W05 &              & \\
GY~310   &      &     & M8.5,M7,M6   & WGM99,LR99,N02 \\
GY~325\tablenotemark{b}   &      &     & M9           & WMG07  \\
GY~350   &      &     & M6           & N02 \\
oph-193  &      &     & M6           & N02 \\
\noalign{\smallskip}
\tableline
\noalign{\smallskip}
\multicolumn{5}{l}{\footnotesize $^a$References same as noted in Table 1, except \citet[][\ H00]{Hillenbrand2000} }\\
\multicolumn{5}{l}{\footnotesize $^b$Needs confirmation as an association member and does not appear in Table 1} \\
\end{tabular}
}
\end{center}
\end{table}

\subsection{Mid-Infrared and Far-Infrared Observations}

The presence or absence of thermal emission by circumstellar
dust is essential in identifying
YSOs and classifying their spectral energy distributions.
While K band observations can often detect inner disk emission,
wavelengths longer than 3.5 $\mu$m are much more sensitive
to circumstellar dust in a disk or envelope.
While the \citet{Wilking1991} review focused on results
from IRAS \citep[e.g.,][]{YoungLadaandWilking1986,WilkingLadaandYoung1989},
there are now mid-infrared cameras available
on ground-based telescopes as well as data from the
Infrared Space Observatory and the Spitzer Space Telescope.

In the post-IRAS period, photometry has been
obtained primarily in the $\lambda$=3.4--24 $\mu$m region for
infrared sources in L1688 that were either too faint or confused
in the IRAS database.  By constructing their
SEDs, or by calculating the spectral indices of their SEDs from
near- to mid-infrared wavelengths, new YSOs can be identified
and their luminosities estimated.
In lieu of detailed knowledge of the disk mass and accretion rate,
the SED class has also been used to infer the evolutionary state.
However,
the viewing angle or foreground extinction could cause YSOs
in different evolutionary states to display
the same spectral index \citep{AndreandMontmerle1994,Robitaille2006}.
With this caveat in mind, the relative numbers of sources in each SED
class is a clue to the duration of each evolutionary stage.
A survey by \citet{Greene1994}, when combined with earlier studies,
derived spectral indices from 2.2 to 10 $\mu$m for 71 YSOs.
In the sample with L $>$ 1 L$_{\sun}$, they found 11 Class I SEDs,
5 flat-spectrum (F) sources, and 22 Class II SEDs.
This implied roughly equal evolutionary timescales
for the embedded states (Class I and F) compared to the CTTS stage
(Class II).  The role of accretion in enhancing the luminosity of
Class I ``objects" relative to Class II
``objects" was evident.

Ground-based mid-infrared studies have an advantage in spatial
resolution over space-borne telescopes
which is especially important given the higher incidence of
multiplicity in YSOs compared to main sequence
stars and the arcsecond scale of circumstellar disks.
Mid-infrared imaging has been used to discover an infrared
companion to the binary CTTS system WL~20 \citep{ResslerandBarsony2001}.
A comprehensive, high resolution (0.25-0.5\arcsec) N-band survey of
172 YSOs in L1688 resolved 15 multiple systems \citep{Barsony2005}.
The survey revealed that there is a
wide range of veiling within a particular SED class suggesting that
accretion is highly time-variable.  A time
scale of $\sim$4 $\times$ 10$^5$ years is derived for
the clearing of the remnant infalling envelope.  A wider survey
of embedded sources in nearby clouds including L1688 has found that
multiple systems with a Class I protostar are most often paired
with a flat-spectrum source and that the envelopes surrounding these objects
must be rapidly evolving \citep{Haisch2006}.

\subsubsection{The Infrared Space Observatory}

Launched in 1995, the Infrared Space Observatory (ISO) provided
improvements in sensitivity and spatial resolution over IRAS.
The entirety of L1688 (and parts of L1689) was surveyed with ISOCAM
at wavelengths of 6.7 $\mu$m and 14.3 $\mu$m.
\citet{Bontempsetal2001} presented photometry for 425 point sources
with a sensitivity down to
10-15 mJy.  Using a spectral index from 2.2 to 14.3 $\mu$m, they found
16 Class I sources, 14 flat-spectrum sources, and 92 Class II sources with
L $>$ 0.03 L$_{\sun}$.  Thus, the ISO survey resulted in a more
complete sampling of objects presumed to be
in the CTTS phase and suggests the embedded state timescale
is roughly one-third of the CTTS stage.  A total of 119 association
members in Table 1 have an infrared excess established from
this survey and are denoted with ``IRX" in the second to last column.
Twenty-three of these YSOs have infrared excesses determined
solely from the ISO survey, primarily due to its more extensive
coverage.
Using the sample of flat-spectrum and Class II objects, \citet{Bontempsetal2001}
constructed the stellar luminosity function for L1688.
They have modeled it with an underlying IMF with a
two-component power law: a high mass index of $-$1.7, a break at about
0.5 M$_{\sun}$, and flat down to 0.06 M$_{\sun}$.  Their result is
consistent with the IMF derived from spectroscopic studies noted
in Section 4.3.

ISOCAM was also used to examine individual objects in L1688.
Using observations at wavelengths of  6.7 $\mu$m and 11.5 $\mu$m,
\citet{Wilkingetal2001} investigated SEDs for 35 radio-emitting
young stars in L1688.  About half of the sample are diskless YSOs
in the high density core with ages comparable to sources with
infrared excesses.  It appears that these objects have shorter disk
survival times compared to CTTSs in the cloud.  The fact that
diskless YSOs are about twice as numerous as CTTSs among
radio-emitting objects emphasizes the essential role that radio
(and X-ray) surveys play in obtaining a complete census of YSOs in
a star-forming region \citep[e.g.,][]{Gagne2004}.

\subsubsection{The Spitzer Space Telescope}

The Spitzer Space Telescope was launched in 2003 and  presented
a hundred-fold increase in sensitivity over ISO.
L1688 has been surveyed by both IRAC and MIPS as part of the
Cores to Disks
(c2d) Legacy progam and Guaranteed Time Observations by the IRAC team.
Fig. 4 shows a three-color IRAC image of a one square parsec area
centered on L1688 obtained by mosaicking data
sets from both programs.  The resulting sensitivity
is 0.015 mJy at 3.6 $\mu$m.  The red color represents emission in the
8 $\mu$m band, green the 4.5 $\mu$m band, and blue the 3.6 $\mu$m band.
Class I sources are enclosed by green circles and are clustered
primarily in molecular cores A (northwest) and E/F (southeast).
The early-type stars Oph Source~1, SR~3, and DoAr~21 are labeled and,
along with the B2V star HD 147889 just off the western edge of the image,
illuminate extensive nebulae of small dust grains.
Fig. 5 is a three-color image of the same area but including MIPS data.
The red color represents emission in the MIPS 24 $\mu$m
band, green the IRAC 8 $\mu$m band, and blue the IRAC 4.5 $\mu$m band.
The Class I sources marked in Fig. 4 are all bright in the
MIPS 24 $\mu$m band.

\begin{figure}
\begin{center}
\includegraphics[height=4.25in]{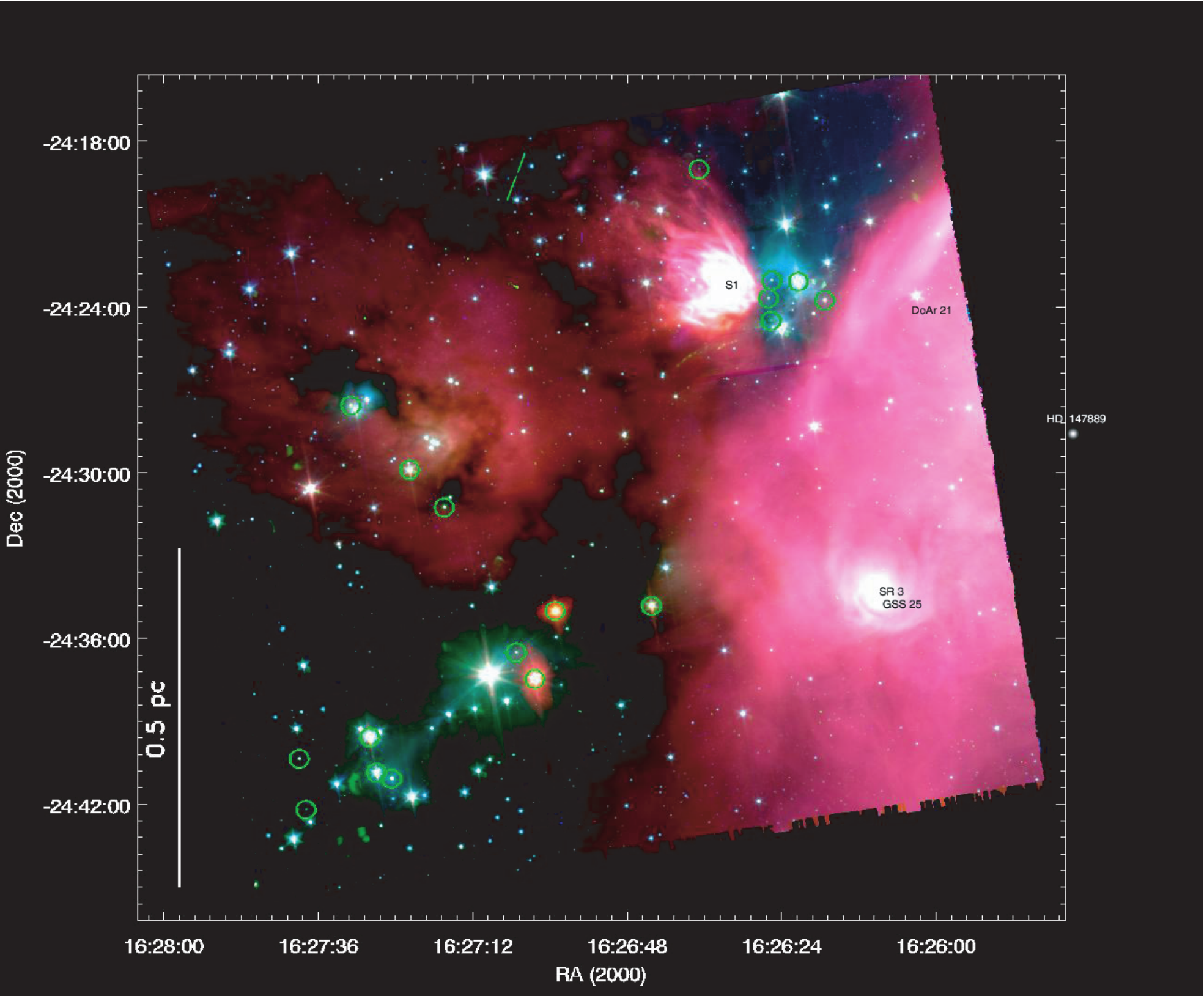}
\end{center}
\caption{The IRAC image of L1688 with red color representing emission in the
8 $\mu$m band, green the 4.5 $\mu$m band, and blue the 3.6 $\mu$m band.
The early-type stars HD~147889, Oph Source~1, SR~3, and DoAr~21 are labeled.
Class I sources are enclosed by green circles.}
\end{figure}

A thorough analysis of the MIPS data covering 14.4 deg$^2$
has been presented by \citet{Padgett2008}.  Their study covered not
only L1688, but much of the dark cloud complex.  A total of 323 candidate
YSOs with infrared excesses were identified from color-magnitude diagrams using the 2MASS K$_s$ band magnitude
and the K$_s$ - MIPS 24 $\mu$m color, and comparisons with off-cloud fields
dominated by extragalactic sources.
One hundred sixty-one of the candidates
reside in L1688, 27 in L1689, and 13 in L1709.  The majority of candidates (84\%) with
Class I or flat-spectrum SEDs are associated with the known YSO aggregates.
However, a significant fraction of
the candidates in the Class II phase (37\%) form a distributed population.

\begin{figure}[h]
\begin{center}
\includegraphics[width=4.75in]{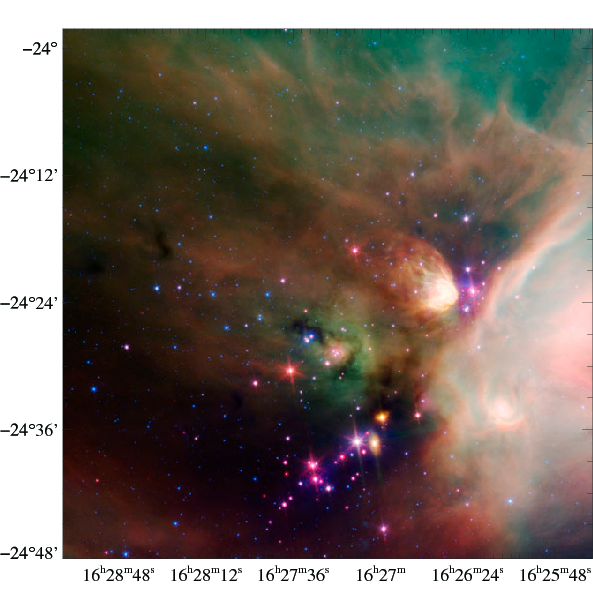}
\end{center}
\caption{A combined IRAC/MIPS image of L1688 with the red color representing
emission in the MIPS 24 $\mu$m
band, green the IRAC 8 $\mu$m band, and blue the IRAC 4.5 $\mu$m band
(image courtesy of Robert Hurt).}
\end{figure}

For this review, Spitzer data for the one square degree area centered on L1688
were combined in three ways to establish
the presence of an infrared excess and association membership. First, there were 116 sources
detected in all four IRAC bands and in the MIPS 24 $\mu$m band.  This
permitted us to calculate the 3.6--24 $\mu$m spectral index {\it a} and
classify 10 objects with Class I SEDs ({\it a}$>$0.3), 31 with flat-spectrum
SEDs ($-$0.3$<${\it a}$<$0.3), 65 with Class II SEDs
($-$1.6$<${\it a}$<$$-$0.3), and 10 with Class III SEDs ({\it a}$<$$-$1.6).
A plot of these sources is shown in an IRAC-MIPS color-color diagram
in Fig. 6.  The sample is incomplete for Class II and Class III objects.
While the spectral index declines continuously from Class I to Class II,
there is a clear break between excess and non-excess sources in this
diagram.  Four of the 106 objects with infrared excesses were not
detected by the 2MASS survey and hence they
do not appear in Table 1.\footnote{The four excess
sources not present in the 2MASS catalog and
Table 1 include possible Class I sources IRAC 162625.61$-$242429
(also the radio continuum source LFAM~4 = GDS J162625.7$-$242429),
and IRAC 162714.3$-$243132 (also the K=15.26 source CRBR/Oph 2412.9$-$2447,
Rieke \& Rieke 1990).  IRAC 162625.97$-$242340 has a flat-spectrum
SED and is associated with a K=17.5 NSFCAM source also observed with HST
as AMD 162625$-$242339 \citet{Allen2002,Wilking2007}.  The
possible Class II source IRAC 162656.42$-$243301.5 has no known
near-infrared or radio counterpart.}
The 102 objects which displayed an infrared
excess based on the 3.6--24 $\mu$m spectral index and which are in the
2MASS catalog do appear in Table 1 along with their SED Class
(I, F, II).   The ``Mex" in the second to last column denotes their identification
as association members based upon the MIPS 24 $\mu$m excess.  Eighteen
of these YSOs have infrared excesses which were not detected by ISO.

\begin{figure}
\begin{center}
\includegraphics[width=4.5in]{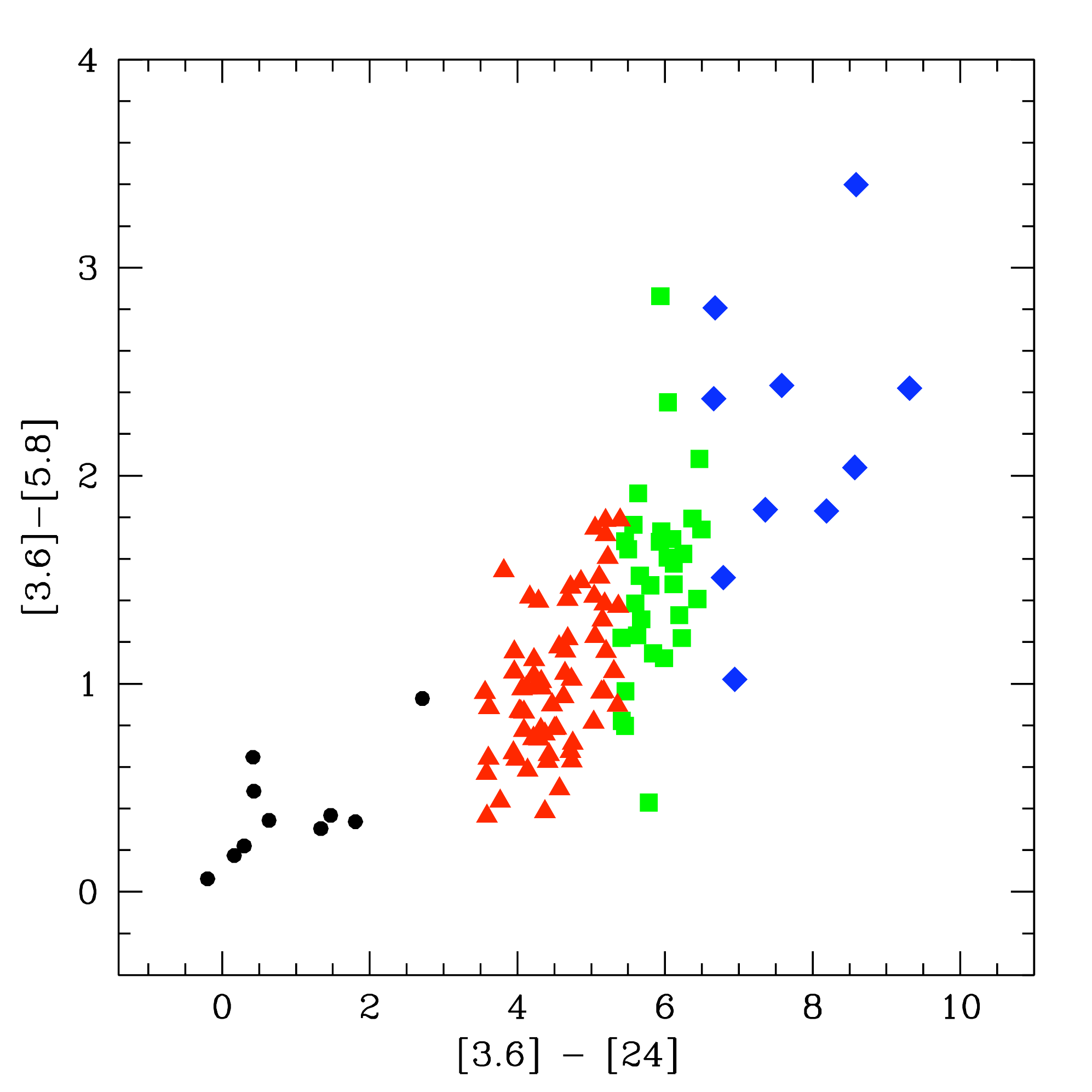}
\end{center}
\caption{A ([3.6]$-$[5.8]) vs. ([3.6]$-$[24]) diagram for 116 sources detected in
all four IRAC bands and in the MIPS 24 $\mu$m band.  Symbols are based on
the spectral index from 3.6--24 $\mu$m: diamonds are Class I sources,
squares are flat-spectrum sources, triangles are Class II sources, and
circles are Class III sources.}
\end{figure}

A second means to determine an infrared excess from the Spitzer data
was to use the more sensitive IRAC data.  Following \citet{Allenetal2004},
a IRAC color-color diagram was constructed for 342 sources with
detections in all 4 bands and errors less than 0.1 mag.  In this
diagram, shown in Fig. 7, Class II objects fall into the region
where the ([3.6]$-$[4.5]) color $<$ 0.8 mag and the ([5.8]$-$[8.0])
color $>$0.2.  Class I objects have ([3.6]$-$[4.5]) colors $>$ 1 mag
while objects with 0.8 $<$ ([3.6]$-$[4.5]) $<$ 1 colors are
associated with flat-spectrum sources.   Consequently,
144 objects fall into the regime of YSOs with infrared excesses:
23 Class I, 28 flat-spectrum, and 93 Class II sources.
A total of 30 new excess sources were identified in this manner.
Clearly, the higher sensitivity of the IRAC bands give a much
more complete sample of excess sources in all evolutionary states.
These sources appear in Table 1 along
with their IRAC classification 1, F, or 2 and a ``B4ex" in the second to last
column.
There are also 38 sources classified as Class III objects
in Fig. 7 that are known association members based on other criteria.
Sources already classified using the
3.6--24 $\mu$m spectral index are shown in Fig. 7 using the same symbols
as Fig. 6. In general, the two methods of classification agree,
although it is clear that reddening can affect the placement of a
source in this diagram.

\begin{figure}
\begin{center}
\includegraphics[width=4.5in]{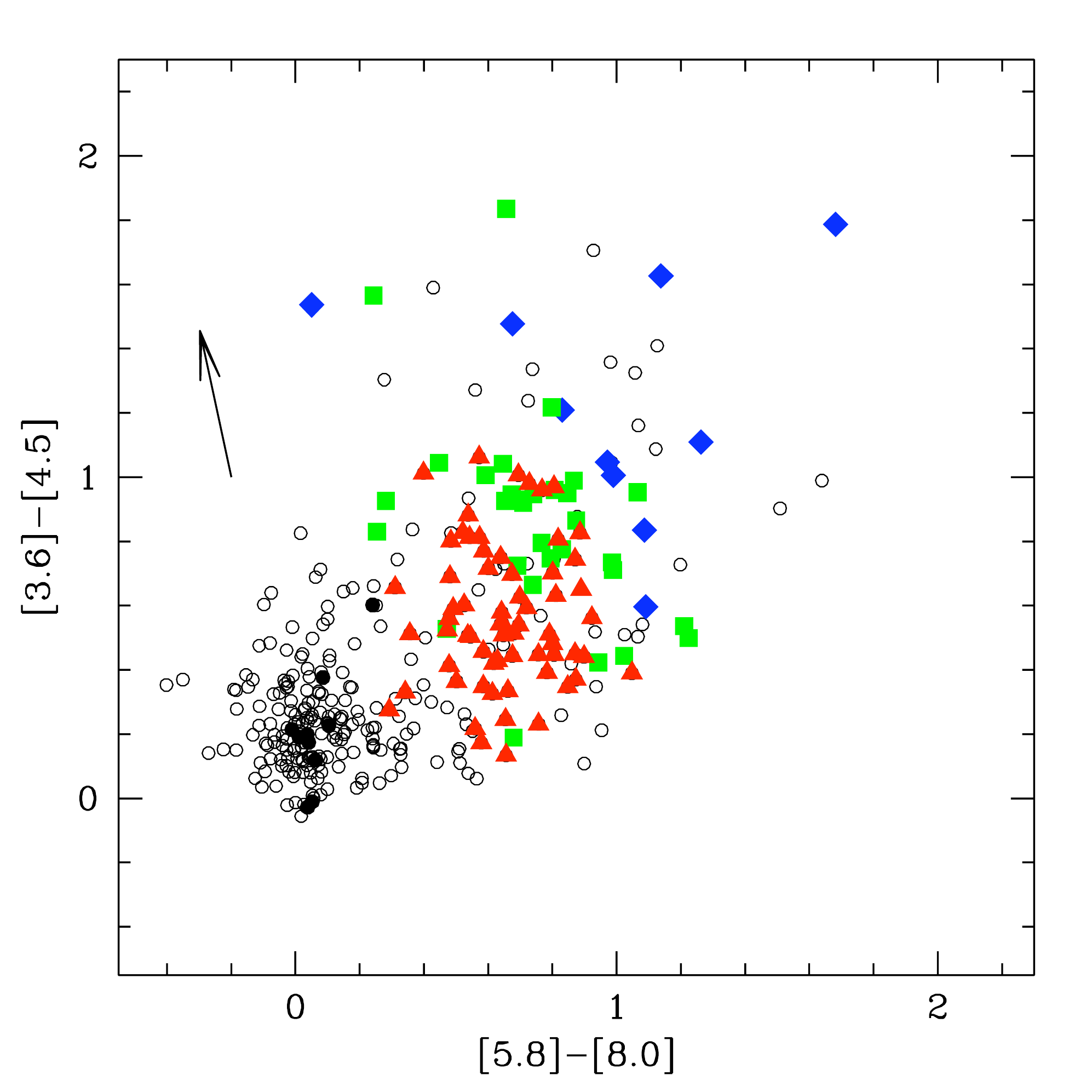}
\end{center}
\caption{A ([3.6]$-$[4.5]) vs. ([5.8]$-$[8.0]) diagram for 342 sources
detected in all four IRAC bands and with errors less than 0.1 mag.
The solid symbols represent the same SED classifications as in Fig. 6
based on the 3.6--24 $\mu$m spectral index.  The arrow marks the
reddening vector for an A$_v$ = 30 mag.}
\end{figure}

A third method used in an attempt to find infrared excesses
for fainter objects
employed a 2MASS-IRAC color-color diagram.
While not shown here, a (H$-$K$_s$) vs. (K$_s$$-$4.5) diagram was
constructed to search for an infrared excess in IRAC band 2.
Assuming a slope for the reddening line of 1.0
\citep{Indebetouw2005}, only seven new excess sources were found
whose photometric errors placed them at least 2 standard deviations
to the red of the reddening line of an M9~V star.  These objects are noted by
``B2ex" in the second to last column of Table 1.

\subsubsection{Mid-Infrared Spectroscopy of Circumstellar Material}

Both absorption and emission from circumstellar gas and dust have been
observed toward YSOs in L1688 and L1689 in an effort to understand more clearly the composition
and strucure of circumstellar disks and
the chemical processes in disks.
Using Spitzer's Infrared Spectrograph (IRS), surveys of T Tauri stars, including objects in L1688
and L1689, have been conducted to search for silicate emission, PAH (polycyclic
aromatic hydrocarbon) emission, as well as gas phase lines \citep{Kessler2006,Geers2006,
Lahuis2007}.  Emission from both amorphous and crystalline silicates
are observed for CTTSs in L1688 and, with the larger sample, follow a trend of
larger silicate grain sizes for cooler YSOs.  PAH emission is relatively
rare in CTTSs, but has been spatially resolved with ground-based imaging for WLY~2-48
and WL~16 (a Herbig Ae star),
and detected but unresolved in SR~21 \citep{ResslerandBarsony2003,Geers2006,Geers2007,Geers2007b}.
The inferred disk diameters are $\sim$100 AU with the inner disk of WLY~2-48 in a transitional
state where it is dominated by small grains.
WLY~2-46 is the first solar mass YSO with strong absorption bands from hot organic molecules,
most likely originating from the inner region of the disk \citep{Lahuis2006} and GSS~30-IRS1
displays rare CO rovibrational emission lines \citep{Dartois2003}.
Using both IRS and MIPS,
the spectral energy distributions of SR~21 (VSSG~23) and 2MASS~J16281370$-$2431391 (the ``Flying Saucer")
have been modeled.
SR~21 has a ``cold disk" with an inner disk depleted of dust \citep{Brown2007,AdamsLadaandShu1987}
suggestive of a planet formation phase.
2MASS~J16281370$-$2431391 is modeled as an edge-on disk system with
the surprising result that large grains ($>$10 $\mu$m) appear not to have settled in the
outer disk \citep{Grossoetal2003,Pontoppidan2007}.

\citet{Alexander2003} used the ISOCAM-CVF from 5 to 16.5 $\mu$m to study
ices in several regions of low mass star formation
including L1688.  Spectra of ten YSOs in cores A and E of the cloud
suggest that L1688 has an underabundance of ice relative to
silicates compared to the R CrA and Serpens clouds.  A larger
scatter in the CO$_2$:H$_2$O ratio was also observed in L1688 compared
to other clouds implying a larger
variation in the local conditions sampled.  Indeed, by focusing on five sources in
core F, \citet{Pontoppidan2006} used ground-based data coupled with data
from ISOCAM and Spitzer's IRS to reveal factor of three variations in this ratio
and the enhanced freeze-out of CO and CO$_2$
toward the center of core F.  Detailed modeling of one source, CRBR 2422.8$-$3423,
suggests that most of the CO ices originate in core material while up to
50\% of the CO$_2$, H$_2$O, and minor ice species reside in the disk \citep{Pontoppidan2005}.

\subsection{Millimeter and Submillimeter Surveys}

Cold dust structures surround YSOs in the earlier phases
of evolution.  Compact dust emission ($<$1\arcsec\ or $<$130 AU)
is thought to arise from a circumstellar disk while
larger structures with radii of 1000-5000 AU delineate an
extended circumstellar envelope \citep[e.g.,][]{Looney2000}.
Single dish mm and submillimeter continuum
surveys, sensitive mainly to the extended envelopes,
have been effective in not only identifying YSOs but also
their evolutionary states
\citep{Andreetal1990b,AndreWard-ThompsonandBarsony1993,AndreandMontmerle1994}.
For example, there is evidence that as a source evolves from
a Class I protostar to a CTTS, a larger fraction of the cold dust
emission resides in the unresolved disk component.  Similarly,
WTTSs (Class III energy distributions) have little or no
cold circumstellar dust \citep[e.g.,][]{AndreandMontmerle1994}.  These
mm continuum surveys have been important in establishing the
Class 0 sources with more massive envelopes such as VLA~1623 and
possibly LFAM~1 and MMS~126
in L1688 and IRAS 16293$-$2422 in L1689.  In the
L1688 core, mm continuum observations have been crucial to
determining the evolutionary state of heavily reddened YSOs.
For example, high extinction can
shift the spectral energy distribution of the CTTSs to resemble
that of less evolved Class I protostars.  However, such objects
will display weaker mm continuum than true Class I protostars.

\addtolength{\tabcolsep}{-0.55mm}

\begin{table}[!hbt]
\vspace{-5mm}
\caption{Embedded Sources in L1688}
\smallskip
\begin{center}
{\small
\begin{tabular}{l@{\hskip8pt}r@{\hskip8pt}c@{\hskip8pt}c@{\hskip8pt}c@{\hskip8pt}l}
\tableline
\noalign{\smallskip}
Name & ISO-Oph & IRAC SED & Sp. Index & MM Cont. & Ref.\tablenotemark{a} \\
     &   No.   &  Class   &3.6-24 $\mu$m & Class &  \\
\noalign{\smallskip}
\tableline
\noalign{\smallskip}

GSS~26       & 17   & 2     & II    &  I?   &   SST,SSGK06 \\
CRBR~12      & 21   & 1     & I     &  II?  &   SST,MAN98  \\
GSS30-IRS~1  & 29   & 1     & I     &  I    &   SST,WLY89,MAN98 \\
LFAM~1       & 31?  &$\cdots$ & I     &  0/I  &   GWAYL94,MAN98 \\
CRBR~25      & 33   & 1     & F     &$\cdots$&   SST \\
GY~21        & 37   & F     & F     &   I?  &   SST,SSGK06 \\
GY~30        &$\cdots$& 1     & I     &$\cdots$&   SST \\
VLA~1623     &$\cdots$&$\cdots$&$\cdots$& 0     &   MAN98  \\
GY~91        & 54   & 1     & I     &I      &   SST,MAN98  \\
WL~12        & 65   & 1     &$\cdots$&  I    &   SST,WLY89,MAN98 \\
GSS~39       & 67   & 2     & II    & I?    &   SST,SSGK06 \\
CRBR~51      & 85   & F     & II    & I     &   SST,MAN98 \\
WL~22        &90    & 1     & I     & I     &   SST,MAN98 \\
WL~16        & 92   & 1     & I     & II    &   SST,WLY89,MAN98  \\
GY~197       & 99  & 1     & I     & I     &   SST,MAN98 \\
GY~201       &$\cdots$& 1     &$\cdots$&$\cdots$&   SST \\
EL~29        &108   & 1     & I     &  I    &   SST,WLY89,MAN98 \\
WL~19        &114   & 1     & II    &  II   &   SST,MAN98 \\
GY~236       &118   & 1     & II    &  I?   &   SST,MAN98 \\
WL~20S       &121?  &$\cdots$& I     & I-II  &   RB01,MAN98 \\
IRAC~161     &$\cdots$& 1   & F     &$\cdots$&   SST \\
IRS~37       &124   & 1     & F     &  II?  &   SST,MAN98  \\
WL~3         &129   & 1     & F     & II    &   SST,MAN98 \\
WL~6         &134   & 1     & F     &  I    &   SST,MAN98 \\
CRBR~85      &137   & 1     &$\cdots$& I     &   SST,MAN98 \\
IRS~43       &141   & 1     & I     &  I    &   SST,WLY89,MAN98 \\
IRS~44       &143   & 1     & I     &  I    &   SST,WLY89,MAN98  \\
IRS~46       &145   & 1     & F     & I     &   SST,MAN98  \\
VSSG~17      &147   & F     & II    &  I?   &   SST,SSGK06 \\
IRAC~130    &150   & 1     & I     &$\cdots$&   SST \\
GY~284       &151   & 2     & F     &  I    &   SST,MAN98 \\
IRS~48       &159   & 2     & I     & I     &   SST,WLY89,MAN98 \\
GY~312       &165   & 1     & I     &$\cdots$ &   SST \\
IRS~51       &167   & 2     & F     &  I    &   SST,MAN98 \\
IRS~54       &182   & F     &  F    &  I    &   SST,MAN98 \\
\noalign{\smallskip}
\tableline
\noalign{\smallskip}
\multicolumn{6}{l}{\parbox{0.95\textwidth}{\footnotesize $^a$References
    are the Spitzer Space Telescope (SST, this study),
    \citet[][SSGK06]{Stankeetal2006},
    \citet[][\ MAN98]{MotteAndreandNeri1998},
    \citet[][\ WLY89]{WilkingLadaandYoung1989},\citet[][\ GWAYL94]{Greene1994},
    and \citet[][\ RB01]{ResslerandBarsony2001}}} \\
\end{tabular}
}
\end{center}
\vspace{-5mm}
\end{table}
\addtolength{\tabcolsep}{0.55mm}

Recent major mm continuum surveys of L1688 have used the IRAM 30-m and
the MPIfR 19-channel bolometer array giving 11\arcsec\ resolution
\citep{MotteAndreandNeri1998}, the JCMT and the SCUBA bolometer array
yielding 14\arcsec\ resolution \citep{Wilson1999,Johnstone2000,Andrews2007},
and the SEST and the
37-channel SIMBA bolometer array with 24\arcsec\ resolution
\citep{Stankeetal2006}.  \citet{Andrews2007} have used a sample of 147 YSOs in the
Ophiuchus complex to derive the properties of their circumstellar disks.  They find their
basic properties are indistinguishable from those in the Taurus-Auriga cloud.  For
Class~II objects, the combined sample yields characteristic values for the disk
mass of $\sim$0.005 M$_{\sun}$ and for the ratio of the disk to stellar mass of $\sim$1\%.
These surveys have also resulted in a complete
census of Class 0 and Class I protostars in L1688.  These results
are summarized in Table 4 and their distribution shown in Fig. 3.  Among these objects, VLA~1623 and
MMS~126 have too much extinction to have detectable
counterparts at $\lambda$=2 $\mu$m in the 2MASS
catalog.  In addition, five sources were detected in the Spitzer
IRAC survey of L1688 with no counterparts in the 2MASS catalog.
However, their nature as Class I protostars has yet to be confirmed.

Interferometric mm continuum observations have targeted individual
sources such as VLA~1623, GSS30/LFAM~1, and EL~29 in L1688 and IRAS
16293$-$2422 in L1689
\citep{Pudritzetal1996,Looney2000,Zhang1997,Boogert2002,Mundyetal1992,Patience2008}.
These are especially critical in identifying Class I and Class 0
binary systems, such as as VLA~1623 and IRAS 16293$-$2422, and
exploring the evolution of circumstellar disks.

\section{Jets and Outflows}

Mass loss activity has been observed by searching for high
velocity CO gas, shock excited clumps of atomic or molecular
gas, radio continuum emission from ionized gas, water masers, and
extended, highly-polarized circumstellar dust.
Since the previous review \citep{Wilking1991}, the number of molecular outflows
in L1688 has increased in number from 4 to 16. Most of these
new outflows have been identified
by observing the higher J transitions of CO \citep{Bontempsetal1996,
Sekimotoetal1997,Kamazaki2001,
Ceccarellietal2002,Kamazakietal2003,Stankeetal2006,Bussmann2007}.  Near-infrared
polarimetry has been used to model emission surrounding WLY~2-54 as an outflow
cavity \citep{Beckford2008}.
The most spectacular outflow remains that from the Class 0 source VLA~1623
\citep{Andreetal1990a,Dentetal1995}
which is also associated with a Herbig-Haro object, numerous H$_2$ emission knots,
and water masers (see below).

In 1991, at the time of the previous review, only two Herbig-Haro (HH)
objects in L1688 had been reported (HH~79 and HH~224).  This was
clearly a selection effect
caused by the high visual extinction in the core.  Although deep optical
imaging surveys of L1688 in H$\alpha$ and [S~II] have now discovered about 33
HH objects (including components of HH objects), they lie preferentially
toward the lower extinction cloud edges and it is often difficult to
identify the exciting star \citep{Wilkingetal1997,GomezWhitneyandWood1998,
Gomezetal2003,PhelpsandBarsony2004}.
Five of these HHs form a jet-like structure that
appears to emanate from the CTTS SR~4.  In addition, a survey of an
11 deg$^2$ region of the Ophiuchus complex has revealed seven groups of HHs
in areas outside of L1688 \citep{Wu2002}.

Because of the high extinction in the L1688 core, searches for shock-excited
clumps using the 1-0 S(1) transition of H$_2$ at 2.12 $\mu$m
have been very successful.  Knots of emission have been
observed in the VLA 1623 outflow \citep{Dentetal1995,DavisandEisloffel1995}.
Large scale surveys of L1688 have identified a total of
74 knots of H$_2$ emission \citep{Grossoetal2001,Gomezetal2003,Khanzadyanetal2004}.
\citet{Khanzadyanetal2004} find that 10 outflows can account for all of the
H$_2$ knots with a tendency for the outflow axes to be perpendicular
to the elongation of the cloud filaments.
They link together H$_2$ emission along with known HH objects
to trace major outflows from the Class I objects YLW 16 (WLY~2-44) and
YLW 15 (WLY~2-43).  \citet{Ybarra06} have detected a series of H$_2$ knots within
30\arcsec\ of EL~29 which trace a precessing jet that has excavated a wide-angle
cavity observed in scattered light.

Radio continuum and water maser emission
have been shown to be effective tracers of mass loss
from YSOs independent of the high visual extinction in the core.
Both tracers can be observed interferometrically and
are either coincident, or within a hundred AUs, of the YSO making
identification of the source of excitation relatively unambiguous.
Radio continuum emission due to free-free emission from ionized gas
has been observed in surveys of the dense cores in L1688
with extended structure associated with jets
\citep{Leousetal1991,Bontemps1996,BontempsandAndre1997,
Gagne2004}.
However, some fraction of the radio continuum sources are extragalactic.
Since the first reports of highly variable maser emission from IRAS 16293$-$2422
in L1689 and YLW 16A (WLY~2-44) and YLW 16B (WLY~2-46) in L1688, ~~water
masers have also been detected toward GSS 30-IRS1 and VLA 1623 in L1688
\citep{Claussenetal1996}.  Subsequently, multi-epoch VLBI observations have
been reported for IRAS 16293$-$2442 \citep{Woottenetal1999,ImaiIwataandMiyoshi1999}
and YLW16A \citep{Simpsonetal2004}
The success of these observations have opened the possibility
of using the VLBA observations to derive accurate parallaxes to the
masers.

\section{Star Formation in the Ophiuchus Streamers}

\begin{figure}
\begin{center}
\includegraphics[width=4.5in]{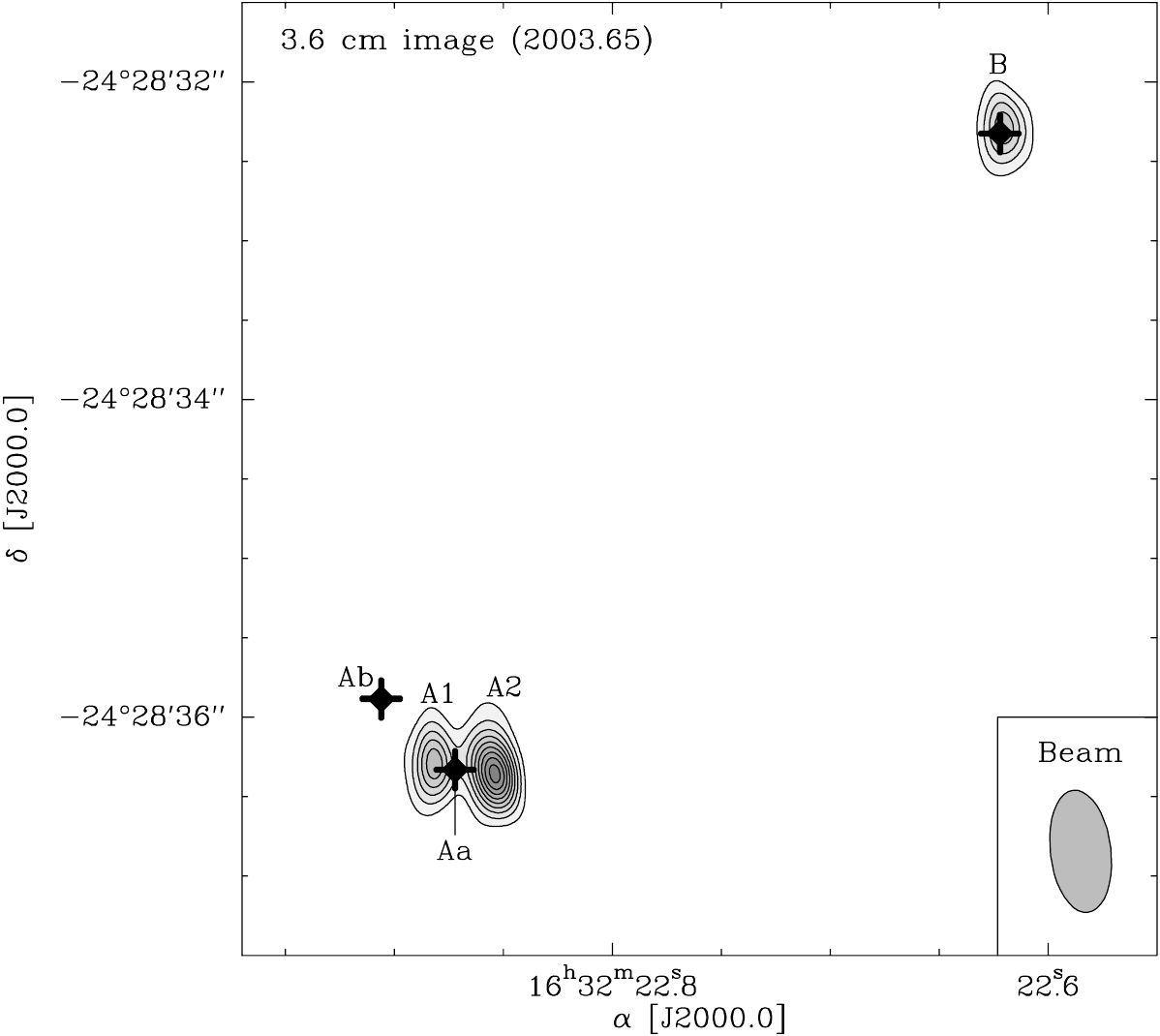}
\end{center}
\caption{A 3.6 cm continuum image of the IRAS 16293$-$2422 region \citep{Loinard2007}.  The
positions of the radio continuum sources A1, A2, and B are shown in relation to the submillimeter
sources Aa, Ab, and B.}
\end{figure}

As shown in Figs. 1 and 2, streamers of molecular gas extend northeast of L1688 (L1709, L1704)
and eastward from the L1689 cloud (L1689, L1712, L1729, see also
Cambr{\' e}sy 1999 and Lombardi, Lada, \& Alves 2008).
These regions have been included in an H$\alpha$ objective prism survey
\citep{WilkingSchwartzandBlackwell1987}, and the IRAS survey.  All of L1689, L1712, L1729,
and L1709 have been included in the 1.1 mm continuum survey by \citet{Young2006} and the MIPS
survey reported by \citet{Padgett2008}.
All of L1689 has been observed at 450 and 850 $\mu$m by \citet{Nutter2006}.
Portions of L1689 were covered by {\it Einstein} X-ray observations
\citep{Montmerle1983}, and ISO \citep{Bontempsetal2001}.
But while comprehensive infrared studies of star formation in the streamers are just beginning,
it is already apparent that there is less high density gas and less star formation compared to L1688
\citep{Nutter2006}.

\begin{figure}
\includegraphics[width=5.25in]{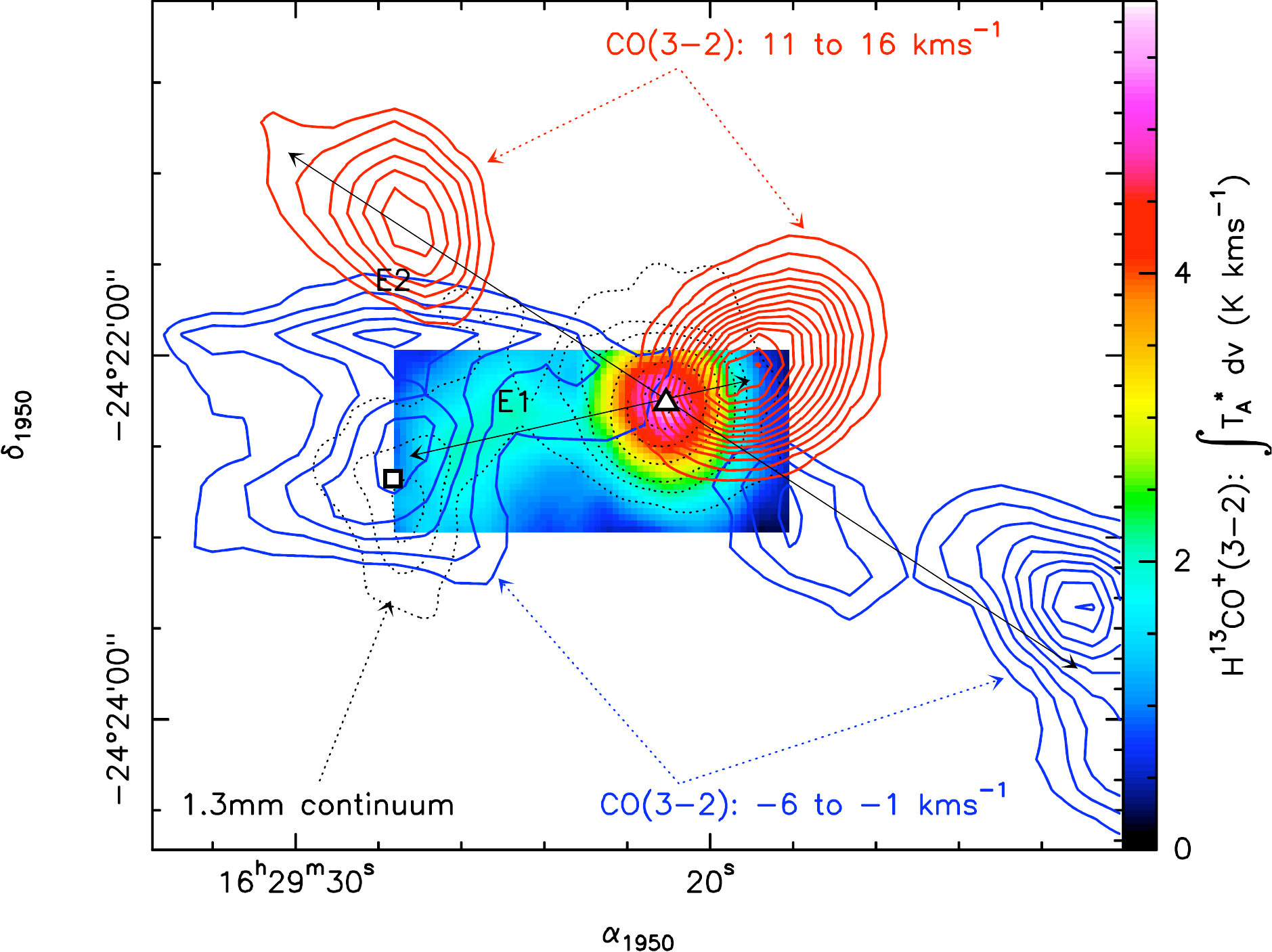}
\caption{A CO(3$-$2) map of the molecular outflows associated with IRAS 16293$-$2422.  Also
shown is the dense gas as traced by H$^{13}$CO$^+$(3$-$2) integrated intensity (color image) and
cold dust traced by 1.3
millimeter continuum emission (black dotted contours) \citep{Lis2002}.
}
\end{figure}

There have been detailed studies of
individual dense cores or YSOs in these regions.
Seven YSOs identified through the IRAS catalog were included in a mid-infrared study by \citet{Greene1994}.
Subsequently two of these YSOs, L1709B/GWAYL~4
and L1689S/GWAYL~6 were observed to have strong mm continuum emission
\citep{AndreandMontmerle1994} and were
observed spectroscopically \citep{Doppmannetal2005}.  Molecular outflows that had been
previously reported for these sources were mapped by \citet{Bontempsetal1996}.

Most of the recent studies in the streamers center around the
binary Class 0 source in L1689, IRAS 16293$-$2422,
and its associated dense gas.
VLBI measurements of the parallax of water masers associated with
IRAS 16293$-$2422 yield a distance of 178 pc (+18 pc, $-$37 pc)
\citep{Imai2007}.
As shown in Fig. 8, the main components of IRAS 16293$-$2422 are separated by 5.2\arcsec\ (900 AU at 175 pc)
and have distinctly different properties:
the southern component, I16293A, is the strongest source of line emission with associated water masers
and free-free emission while the northern component, I16293B, shows compact emission dominated by
optically thick dust from 5-300 GHz
\citep{Mundyetal1992,Wootten1993,Rodriguez2005,Chandler2005,Remijan2006,Loinard2007}.

Recent studies of I16293A by \citet{Chandler2005} have resolved the southern source into
two components separated by 0.64\arcsec\ (Aa and Ab) which are distinct from the centimeter sources
A1 and A2.  On the larger scale, IRAS 16293$-$2422 appears to drive
two distinct bipolar CO outflows first mapped by \citet{WoottenandLoren1987},
\citet{Walkeretal1988}, and \citet{Mizuno1990}.  More recently, the outflows have been studied at higher resolution and
in different molecules \citep{Castets2001,Hirano2001,Garay2002,Lis2002,Starketal2004,Yeh2007}.  Fig. 9
shows the molecular outflow in relation to the dense core.

It has been suggested that the northeast-southwest outflow is driven by source A2 in
I16293A and the east-west outflow originates from
I16293B \citep{Chandler2005,Loinard2007}.
A rich spectrum of complex organic molecules is observed from compact ($<$1.5\arcsec) regions
toward I16293A and I16293B
\citep{Cazaux2003,Bottinelli2004,Schoier2004,Kuan2004}.
\citet{Cazaux2003} have suggested that the distribution of molecular emission is consistent with
grain evaporation in the near-circumstellar environment.  This has been called into question
by \citet{Jorgensen2005} who find evidence for an inner cavity in the protobinary system with
a radius of 600 AU.  Indeed, \citet{Chandler2005} show that the molecular emission toward
I16293A coincides with centimeter source A1 and is likely to be shock-excited by the outflow from A2.

\acknowledgments
We thank Robert Gendler for permission to use Figure 1.\linebreak
Naomi Ridge generously made the COMPLETE data of the Ophiuchus molecular
complex in Fig. 2 available and Di Li allowed us to display his C$^{18}$O map of L1688 in advance of
publication.  We also thank Robert Hurt for preparing the color image shown in Figure 5.
This review has made extensive use of NASA's Astrophysics Data System, and
the SIMBAD database,
operated at CDS, Strasbourg, France.  This publication also makes use of data products from the
Two Micron All Sky Survey, which is a joint project of the University of Massachusetts
and the Infrared Processing and Analysis
Center/California Institute of Technology, funded by the National Aeronautics and Space
Administration and the National Science Foundation.

{\bf Appendix A: Description of Table 1}

A list of association members in the L1688 cloud is presented in
Table~1 (see footnote in Section~4 for electronic access to Table~1).
Selection of objects for Table 1 is described in Section~4.
All of the objects listed in Table 1 appear in the 2MASS catalog with,
at minimum, a detection in the K$_s$ band.  In addition to their
positions and J, H, and K$_s$ magnitudes from the 2MASS catalog
(col. 1-13), we list source names from selected surveys including
H$\alpha$ surveys of \citet[][\ SR]{StruveandRudkjobing1949} and
\citet[][\ WSB]{WilkingSchwartzandBlackwell1987}; near-infrared
surveys of \citet[][\ VSSG]{Vrbaetal1975} and \citet[][\
GY]{GreeneandYoung1992}; and far-infrared surveys of \citet[][\
ISO-Oph] {Bontempsetal2001} and \citet[][\ IRAC]{Allenetal2006}.
Names listed under ``Other" (column 21) include selected
identifications used in the literature from the H$\alpha$ survey of
\citet[][\ DoAr]{DolidzeandArakelyan1959}; the optical surveys of
\citet[][\ C]{Chini1981}, \citet[][\ ROX]{BouvierandAppenzeller1992}
\citet[][\ MMGC]{Martinetal1998}; the infrared surveys of \citet[][\
GSS or Source] {GrasdalenStromandStrom1973}, \citet[][\
VSS~II]{Vrbaetal1976}, \citet[][\ EL]{Elias1978}, \citet[][\
WL]{WilkingandLada1983}, \citet[][\ WLY]{WilkingLadaandYoung1989},
\citet[][\ CRBR]{Comeronetal1993}, \citet[][\
SKS]{StromKepnerStrom1995}, or \citet[][\ AMD]{Allen2002}; or the
radio continuum survey of \citet[][\ LFAM]{Leousetal1991}; ``FS"
refers to the YSO with an edge-on circumstellar disk discovered by
\citet{Grossoetal2003} dubbed the ``Flying Saucer".  Columns 22 and 23
list the X-ray identifications from the Einstein survey by \citet[][\
ROX]{Montmerle1983}; the ROSAT survey by \citet[][\
ROXR1]{Casanova1995}, \citet[][\ ROXRA or ROXRF]{Grosso2000},
\citet[][\ RXJ]{Martinetal1998}; the ASCA survey by \citet[][\
ROXA]{Kamata1997}; the Chandra survey by \citet[][\
IKT2001]{Imanishi2001}, \citet[][\ GDS]{Gagne2004}, or N. Grosso
(2005, ROXC, private communication); or observations with XMM-Newton
by \citet[][\ XMM]{Bouy2004} or \citet[][\ ROXN]{Ozawa2005}.
Preference was given to sources observed by the higher resolution
ROSAT or CHANDRA telescopes.  SED classifications are given in columns
24 and 25 for IRAC and IRAC/MIPS data from the Spitzer Space
Telescope.  For the latter, Class I sources (I) have a spectral index
from 3.6 to 24 $\mu$m of {\it a}$>$0.3, flat-spectrum sources (F) have
-0.3$<${\it a}$<$0.3, and Class II sources (II) have -1.6$<${\it
a}$<$-0.3.  Spectral types are given in columns 26 and 27 for optical
spectra from the studies of \citet[][\ CK79]{CohenandKuhi1979},
\citet[][\ BA92]{BouvierandAppenzeller1992}, \citet[][\
W94]{Walteretal1994}, \citet[][\ LLR97]{LuhmanLiebertandRieke1997},
\citet[][\ MMGC]{Martinetal1998}, \citet[][\ W05]{Wilking2005},
\citet[][\ S06]{Slesnick2006}, or \citet[][\ G07]{Geers2007}, and for
infrared spectra from the studies of \citet[][\
GM95]{GreeneandMeyer1995}, \citet[][\ GL97]{GreeneandLada1997},
\citet[][\ WGM99]{WilkingGreeneandMeyer1999}, \citet[][\
LR99]{LuhmanandRieke1999}, \citet[][\ CTK00]{Cushing2000}, \citet[][\
N02]{Nattaetal2002}, or \citet[][\ WMG07]{Wilking2007}.  Criteria met
for association membership are listed in column 29 (see text for
explanation of symbols).  The last column notes whether the source has
companions within 10\arcsec\ (1300 AU) with ``B" indicating a binary,
``T" a triple, and ``SB" a spectroscopic binary, followed by the
reference.  When both members of a wide binary appear as separate
entries in the table, no note appears in column 29 (e.g., SR~24N and
SR~24S).  References for multiplicity include \citet[][\
G93]{Ghez1993}, \citet[][\ HM95]{HaffnerandMeyer1995}, \citet[][\
B03]{BarsonyKoreskoandMatthews2003}, \citet[][\ A97]{Ageorges1997},
\citet[][\ RZ93]{ReipurthandZinnecker1993}, \citet[][\
A02]{Allen2002}, \citet[][\ C88]{Chelli1988}, \citet[][\
D04]{Duchene2004}, \citet[][\ R94]{Richichietal1994}, \citet[][\
B89]{Barsony1989}, \citet[][\ C00]{Costa2000}, \citet[][\
S95]{Simonetal1995}, \citet[][\ S87]{Simonetal1987}, \citet[][\
H02]{Haischetal2002}, and \citet[][\ T01]{Terebey2001}.

\begin{thebibliography}{}

\bibitem[Abergel et al.(1996)]{Abergeletal1996} Abergel, A., Bernard, J. P., Boulanger, F.,
               Cesarsky, C., Desert, F. X. et al. 1996, \aap, 315, L329
\bibitem[Adams, Lada, \& Shu(1987)] {AdamsLadaandShu1987} Adams, F. C., Lada, C. J.,
        \& Shu, F. H. 1987, \apj, 312, 788
\bibitem[Ageorges et al.(1997)]{Ageorges1997} Ageorges, N., Eckart, A., Monin, J.-L.,
       \& Menard, F. 1997, \aap, 326, 632
\bibitem[Alexander et al.(2003)]{Alexander2003} Alexander, R. D., Casali, M. M., Andr{\' e}, Ph., Persi, P.,
       \& Eiroa, C. 2003, \aap, 401, 613
\bibitem[Allen et al.(2002)]{Allen2002} Allen, L. E., Myers, P. C., Di Francesco, J., Mathieu, R., Chen, H.,
       \& Young, E. 2002, \apj, 56, 993
\bibitem[Allen et al.(2004)]{Allenetal2004} Allen, L. E., Calvet, N., D'Alessio, P.,
           Merin, B., Hartmann, L., et al. 2004, \apjs, 154, 363
\bibitem[Allen et al.(2008)]{Allenetal2006} Allen, L. E., Megeath, T., Padgett, D., Wilking, B., Gagn{\' e}, M., et
          al. 2008, in preparation
\bibitem[Alves de Oliveira \& Casali(2008)]{Alves2008} Alves de Oliveira, C. \& Casali, M. 2008, \aap, 485, 155
\bibitem[Andr{\' e} et al.(1990a)]{Andreetal1990a} Andr{\' e}, Ph., Mart{\' i}n-Pintado, Depois, D.,
       \& Montmerle, T. 1990a, \aap, 236, 180
\bibitem[Andr{\' e} et al.(1990b)]{Andreetal1990b} Andr{\' e}, Ph., Montmerle, T., Feigelson, E. D.,
       \& Steppe, H. 1990b, \aap, 240, 321
\bibitem[Andr{\' e} et al.(1991)]{Andreetal1991} Andr{\' e}, Ph., Phillips, R. B., Lestrade, J.-F.,
        \& Klein, K.-L. 1991, \apj, 376, 630
\bibitem[Andr{\' e} et al.(1992)]{Andreetal1992} Andr{\' e}, Ph., Deeney, B. D., Phillips, R. B.,
        \& Lestrade, J.-F. 1992, \apj, 401, 667
\bibitem[Andr{\' e}, Ward-Thompson, \& Barsony(1993)]{AndreWard-ThompsonandBarsony1993} Andr{\' e}, Ph. Ward-Thompson,
         D., \& Barsony, M. 1993, \apj, 406, 122
\bibitem[Andr{\' e} \& Montmerle(1994)]{AndreandMontmerle1994} Andr{\' e}, Ph.
        \& Montmerle, T. 1994, \apj, 420, 837
\bibitem[Andr{\' e} et al.(2007)]{Andre2007} Andr{\' e}, Ph., Belloche, A., Motte, F.,
        \& Peretto, N. 2007, \aap, 472, 519
\bibitem[Andrews \& Williams(2007)] {Andrews2007} Andrews, S. M. \& Williams, J. P. 2007, \apj, 671, 1800
\bibitem[Barsony et al.(1989)]{Barsony1989} Barsony, M., Burton, M. G., Russell, A. P. G., Carlstrom, J. E.,
        \& Garden, R. 1989, \apj, 346, L93
\bibitem[Barsony et al.(1997)]{Barsonyetal1997} Barsony, M., Kenyon, S., Lada, E.,
        \& Teuben, P. 1997, \apjs, 112, 109
\bibitem[Barsony, Koresko, \& Matthews(2003)]{BarsonyKoreskoandMatthews2003} Barsony, M., Koresko, C.,
        \& Matthews, K. 2003, 591, 1064
\bibitem[Barsony, Ressler, \& Marsh (2005)]{Barsony2005} Barsony, M., Ressler, M.,
        \& Marsh, K. 2005, \apj, 630, 381
\bibitem[Beckford et al.(2008)]{Beckford2008} Beckford, A. F., Lucas, P. W., Chrysostomou,
         \& Gledhill, T. M. 2008, \mnras, 384, 907
\bibitem[Bertiau(1958)]{Bertiau1958} Bertiau, F. C. 1958, \apj, 128, 533
\bibitem[Blaauw(1961)]{Blauuw1961} Blaauw, A. 1961, Bull. Astron. Inst. Neth. 15, 265
\bibitem[Bontemps(1996)]{Bontemps1996} Bontemps, S. 1996, Ph.D. thesis, Univ. Paris, XI
\bibitem[Bontemps et al.(1996)]{Bontempsetal1996} Bontemps, S., Andr{\' e}, Ph., Terebey, S.,
        \& Cabrit, S. 1996, \aap, 311, 858
\bibitem[Bontemps \&  Andr{\' e}(1997)]{BontempsandAndre1997}  Bontemps, S. \& Andr{\' e}, Ph. 1997,
        in {\it Low Mass Star Formation - from Infall to Outflow, Poster proceedings of IAU Symposium
         No. 182}, ed. F. Malbet and A. Castets, 63
\bibitem[Bontemps et al.(2001)]{Bontempsetal2001} Bontemps, S., Andr{\' e}, Ph., Kaas, A. A., Nordh, L.,
         Olofsson, G., et al. 2001, \aap, 372, 173
\bibitem[Boogert et al.(2002)]{Boogert2002} Boogert, A. C. A., Hogerheijde, M. R.,  Ceccarelli, C., Tielens, A. G. G. M.,
        van Dishoeck, E. F., Blake, G. A., Latter, W. B., \& Motte, F. 2002, \apj, 570, 708
\bibitem[Bottinelli et al.(2004)]{Bottinelli2004} Bottinelli, S., Ceccarelli, C., Neri, R., Williams,
        J. P., Caux, E., Cazaux, S., Lefloch, B., Maret, S., \& Tielens, A. G. G. M. 2004, \apj, 617, L69
\bibitem[Bouvier \& Appenzeller(1992)]{BouvierandAppenzeller1992} Bouvier, J.
       \& Appenzeller, I. 1992, \aaps, 92, 481
\bibitem[Bouy, H.(2004)]{Bouy2004} Bouy, H. 2004, \aap, 424, 619
\bibitem[Brandner et al.(1996)]{Brandner1996} Brandner, W., Alcala, J. M., Kunkel, M.,  Moneti, A.,
       \& Zinnecker, H. 1996, \aap, 307, 121
\bibitem[Brown et al.(2007)] {Brown2007} Brown, J. M., Blake, G. A., Dullemond, C. P., Merin, B.,
        Augereau, J. C.  et al. 2007, \apj, 664, L107
\bibitem[Bussmann et al.(2007)] {Bussmann2007} Bussman, R. S., Wong, T. W., Hedden, A. S., Kulesa, C. A.,
       \& Walker, C. K. 2007, \apj, 657, L33
\bibitem[Cambr{\' e}sy(1999)]{Cambresy1999} Cambr{\' e}sy, L., 1999, \aap, 345, 965
\bibitem[Casali \& Matthews(1992)]{CasaliandMatthews1992} Casali, M. M. \& Matthews, H. E. 1992, \mnras, 258, 399
\bibitem[Casanova et al.(1995)]{Casanova1995} Casanova, S., Montmerle, T.,
         Feigelson, E.~D., \& Andr{\' e}, Ph. 1995, \apj, 439, 752
\bibitem[Castets et al.(2001)]{Castets2001} Castets, A., Ceccarelli, C., Loinard, L., Caux, E.,
       \& Lefloch, B. 2001, \aap, 375, 40
\bibitem[Cazaux et al.(2003)]{Cazaux2003} Cazaux, S., Tielens, A. G. G. M., Ceccarelli, C., Castets, A., Wakelam, V.,
        Caux, E., Parise, B., \& Teyssier, D. 2003, \apj, 539, L51
\bibitem[Ceccarelli et al.(2002)]{Ceccarellietal2002} Ceccarelli, C., Boogert, A. C. A., Tielens, A. G. G. M.,
        Caux, E.,  Hogerheijde, M. R., \& Parise, B. 2002, \aap, 395, 863
\bibitem[Chandler et al.(2005)] {Chandler2005} Chandler, C.J., Brogan, C. L., Shirley, Y. L.,
       \& Loinard, L. 2005, \apj, 632, 371
\bibitem[Chelli et al.(1988)]{Chelli1988} Chelli, A., Cruz-Gonzales, I., Zinnecker, H., Carrasco, L.,
       \& Perrier, C. 1988, \aap, 207, 46
\bibitem[Chini(1981)]{Chini1981} Chini, R. 1981, \aap, 99, 346
\bibitem[Claussen et al.(1996)]{Claussenetal1996} Claussen, M. J., Wilking, B. A., Benson, P. J., Wootten, A.,
        Myers, P. C., \& Terebey, S. 1996, \apjs, 106, 111
\bibitem[Cohen \& Kuhi(1979)]{CohenandKuhi1979} Cohen, M. \& Kuhi, L. V. 1979, \apjs, 41, 743
\bibitem[Comer\'on et al.(1993)]{Comeronetal1993} Comer\'on, F., Rieke, G. H., Burrows, A.,
       \& Rieke, M. J. 1993, \apj, 416, 185
\bibitem[Comer\'on et al.(1998)]{Comeronetal1998} Comer\'on, F., Rieke, G. H., Claes, P., Torra, J.,
       \& Laureijis, R. J. 1998, \aap, 335, 522
\bibitem[Costa et al.(2000)]{Costa2000} Costa, A., Jessop, N. E., Yun, J. L., Santos, C., Ward-Thompson, D.,
       \& Casali, M. M. 2000, in Poster Proc. of IAU Symposium 200, {\it Birth and Evolution of Binary Stars},
        ed. B. Reipurth \& H. Zinnecker, 48
\bibitem[Covey et al.(2005)]{Coveyetal2005}  Covey, K. R., Greene, T. P., Doppmann, G. W.,
        \& Lada, C. J. 2005, \aj, 129, 2765
\bibitem[Cushing, Tokunaga, \& Kobayashi(2000)]{Cushing2000} Cushing, M. C., Tokunaga, A. T.,
        \& Kobayashi, N. 2000, \aj, 119, 3019
\bibitem[Cutri et al.(2003)]{Cutri2003} Cutri, R. M., Skrutskie, M. F., van Dyk, S.,
              Beichman, C. A., Carpenter, J. M., et al. 2003, {\it 2MASS All-Sky Catalog
        of Point Sources} (Pasadena: IPAC)
\bibitem[D'Antona \& Mazzitelli(1997)]{DAntonaandMazzitelli1997} D'Antona, F. \& Mazzitelli, I. 1997,
        Mem. S. A. It., 68, 4
\bibitem[Dartois et al.(2003)] {Dartois2003} Dartois, E., d'Hendecourt, L., Thi, W.-F., Pontopiddan, K., Schiutte, W.,
        \& van Dishoeck, E. F. 2004, in {\it Galactic Star Formation Across the Stellar Mass Spectrum},
         ASP Conference Series,
        Vol. 87, eds. J. M. De Buizer \& N. S. van der Bliek, 187
\bibitem[Davis \& Eisl\"{o}ffel(1995)]{DavisandEisloffel1995} Davis, C. J. \& Eisl\"{o}ffel, J. 1995, \aap, 300, 851
\bibitem[de Geus(1992)]{deGeus1992} de Geus, E. J. 1992, \aap, 262, 258
\bibitem[de Geus, de Zeeuw, \& Lub(1989)]{deGeusetal1989} de Geus, E. J., de Zeeuw, P. T.,
       \& Lub, J. 1989, \aap, 216, 44
\bibitem[Dent, Matthews, \& Walther(1995)]{Dentetal1995} Dent, W. R. F., Matthews, H. E.
       \& Walther, D. 1995, \mnras, 277, 193
\bibitem[de Zeeuw et al.(1999)]{deZeeuw1999} de Zeeuw, P. T., Hoogerwerf, R.,
        de Bruijne, J. H. J., Brown, A. G. A., \& Blaauw, A. 1999, \aj, 117, 354
\bibitem[Di Francesco, Andr{\' e}, \& Myers(2004)]{DiFrancescoAndreandMyers2004} Di Francesco, J.,
         Andr{\' e}, Ph., \& Myers, P. C. 2004, \apj, 617, 425
\bibitem[Dolidze \& Arakelyan(1959)]{DolidzeandArakelyan1959} Dolidze, M. V.
        \& Arakelyan, M. A. 1959, Soviet Astr., 3, 434
\bibitem[Doppmann, Jaffe, \& White(2003)]{Doppmann2003} Doppmann, G. W., Jaffe, D. T.
        \& White, R. J. 2003, \aj, 126, 3043
\bibitem[Doppmann et al.(2005)]{Doppmannetal2005}  Doppmann, G. W., Greene, T. P., Covey, K. R.,
        \& Lada, C. J. 2005, \aj, 130, 1145
\bibitem[Duch\^{e}ne et al.(2004)]{Duchene2004} Duch\^{e}ne, G., Bouvier, J., Bontemps, S., Andr{\' e}, Ph.,
        \& Motte, F. 2004, \aap, 427, 651
\bibitem[Duch\^{e}ne et al.(2007)] {Duchene2007} Duch\^{e}ne, G., Bontemps, S., Bouvier, J., Andr{\' e}, Ph.,
        Djupvik, A. A., \& Ghez, A. M. 2007, \aap, 476, 229
\bibitem[Elias(1978)]{Elias1978} Elias, J. H. 1978, \apj, 224, 453
\bibitem[Favata et al.(2005)]{Favata2005} Favata, F., Micela, G., Silva, B., Sciortino, S.,
        \& Tsujimoto, M. 2005, \aap, 433, 1047
\bibitem[Feigelson et al.(2005)]{Feigelson2005} Feigelson, E. D., Getman, K.,
               Townsley, L., Garmire, G.,
               Preibisch, T., et al.
       2005, ApJS, 160, 379
\bibitem[Gagn{\' e} et al.(2004)]{Gagne2004} Gagn{\' e}, M., Skinner, S.~L., \& Daniel, K.~J.
        2004, \apj, 613, 393
\bibitem[Garay et al.(2002)]{Garay2002} Garay, G., Mardones, D., Caselli, P.,
       \& Bourke, T. 2002, \apj, 567, 980
\bibitem[Gatti et al.(2006)]{Gatti2006} Gatti, T., Testi, L., Natta, A., Randich, S.,
       \& Muzerolle, J. 2006, \aap, 460, 547
\bibitem[Geers et al.(2006)] {Geers2006} Geers, V. C., Augereau, J.-C., Pontoppidan, K. M., Dullemond, C. P.,
        Visser, R. et al. 2006, \aap, 459, 545
\bibitem[Geers et al.(2007a)] {Geers2007} Geers, V. C., Pontoppidan, K. M., van Dishoeck, E. F., Dullemond, C. F.,
        Augereau, J.-C., Merin, B., Oliveira, I., \& Pel, J. W. 2007a, \aap, 469, L35
\bibitem[Geers et al.(2007b)] {Geers2007b} Geers, V. C., van Dishoeck, E. F., Visser, R.,  Pontoppidan, K. M.,
          Augereau, J.-C., Habart, E., Lagrange, A. M. 2007b, \aap, 476, 279
\bibitem[Ghez, Neugebauer, \& Matthews(1993)]{Ghez1993} Ghez, A. M., Neugebauer, G.,
       \& Matthews, K. 1993, \aj, 106, 2005
\bibitem[Giardino et al.(2007)]{Giardino2007} Giardino, G., Favata, F., Pillitteri, I., Flaccomio, E., Micela, M.,
       \& Sciortino, S. 2007, \aap, 475, 891
\bibitem[G{\' o}mez, Whitney, and Wood(1998)]{GomezWhitneyandWood1998} G{\' o}mez, M., Whitney, B. A.,
       \& Wood, K. 1998, \aj, 115, 2018
\bibitem[G{\' o}mez et al.(2003)]{Gomezetal2003} G{\' o}mez, M., Stark, D, P., Whitney, B. A.,
       \& Churchwell, E. 2003, \aj, 126, 863
\bibitem[Grasdalen, Strom, \& Strom(1973)]{GrasdalenStromandStrom1973} Grasdalen, G. L. Strom, K. M.,
        \& Strom, S. E. 1973, \apj, 184, L53
\bibitem[Greene \& Lada(1996)]{GreeneandLada1996} Greene, T. P. \& Lada, C. J.
        1996, \aj, 112, 2184
\bibitem[Greene \& Lada(1997)]{GreeneandLada1997} Greene, T. P. \& Lada, C. J.
        1997, \aj, 114, 2157
\bibitem[Greene \& Lada(2000)]{GreeneandLada2000} Greene, T. P. \& Lada, C. J.
        2000, \aj, 120, 430
\bibitem[Greene \& Lada(2002)]{GreeneandLada2002} Greene, T. P. \& Lada, C. J.
        2002, \aj, 124, 2185
\bibitem[Greene \& Meyer(1995)]{GreeneandMeyer1995} Greene, T. P. \& Meyer, M. R.
        1995, \apj, 450, 233
\bibitem[Greene et al.(1994)]{Greene1994} Greene, T. P., Wilking, B. A., Andr{\' e}, Ph., Young, E.,
        \& Lada, C. J. 1994, \apj, 434, 614
\bibitem[Greene \& Young(1992)]{GreeneandYoung1992} Greene, T. P. \& Young, E. T. 1992, \apj, 395, 516
\bibitem[Grosso et al.(2000)]{Grosso2000} Grosso, N., Montmerle, T., Bontemps, S.,
        Andr{\' e}, Ph., \& Feigelson, E. D. 2000, \aap, 359, 113
\bibitem[Grosso et al.(2001)]{Grossoetal2001} Grosso, N., Alves, J., Neuh\"{a}user, R.
      \& Montmerle, T. 2001, \aap, 380, L1
\bibitem[Grosso et al.(2003)]{Grossoetal2003} Grosso, N., Alves, J., Wood, K., Neuh\"{a}user, R.,
        Montmerle, T., \& Bjorkman, J. E. 2003, \apj, 586, 296
\bibitem[Haisch et al.(2002)]{Haischetal2002} Haisch, K., Barsony, M., Greene, T. P.,
       \& Ressler, M. E. 2002, \aj, 124, 2841
\bibitem[Haisch et al.(2004)]{Haischetal2004} Haisch, K., Greene, T. P., Barsony, M.,
       \& Stahler, S. W. 2004, \aj, 127, 1747
\bibitem[Haisch et al.(2006)]{Haisch2006} Haisch, K., Barsony, M., Ressler, M.,
       \& Greene, T. 2006, \aj, 132, 2675
\bibitem[Haffner \& Meyer(1995)]{HaffnerandMeyer1995} Haffner, L. M. \& Meyer, D. M. 1995, \apj, 453, 450
\bibitem[Hillenbrand(1997)]{Hillenbrand1997} Hillenbrand, L. A. 1997, \aj, 113, 1733
\bibitem[Hillenbrand(2000)]{Hillenbrand2000} Hillenbrand, L. A. 2000, unpublished data
\bibitem[Hirano et al.(2001)]{Hirano2001} Hirano, N., Mikami, H., Umemoto, T., Yamamoto, S.,
       \& Taniguchi, Y. 2001, \apj, 547, 899
\bibitem[Imai, Iwata, \& Miyoshi(1999)]{ImaiIwataandMiyoshi1999} Imai, H., Iwata, T.
       \& Miyoshi 1999, \pasj, 51, 473
\bibitem[Imai et al.(2007)] {Imai2007} Imai, H., Nakashima, K., Bushimata, T., Choi, Y. K.,
         Hirota, T. et al. 2007, \pasj, 59, 1107
\bibitem[Imanishi et al.(2001)]{Imanishi2001} Imanishi, K., Koyama, K., \& Tsuboi, Y.
        2001, \apj, 557, 747
\bibitem[Imanishi et al.(2002)]{Imanishi2002} Imanishi, K., Tsujimoto, M.,
       \& Koyama, K.\ 2002, \apj, 572, 300
\bibitem[Imanishi et al.(2003)]{Imanishi2003} Imanishi, K., Nakajima, H.,
        Tsujimoto, M., Koyama, K., \& Tsuboi, Y. 2003, \pasj, 55, 653
\bibitem[Indebetouw et al.(2005)]{Indebetouw2005} Indebetouw, R., Mathis, J. S., Babler, B. L.,
               Meade, M. R., Watson, C., et al. 2005, \apj, 619, 931
\bibitem[Johnstone et al.(2000)]{Johnstone2000} Johnstone, D., Wilson, C. D., Moriarty-Schieven, G., Joncas, G.,
        Smith, G., Gregersen, E., \& Fich, M. 2000, \apj, 545, 327
\bibitem[Jorgensen et al.(2005)]{Jorgensen2005} Jorgensen, J. K., Lahuis, F., Schoeier, F. L.,
              van Dishoeck, E. F., Blake, G. A., et al. 2005, \apj, 631, l77
\bibitem[Kamata et al.(1997)]{Kamata1997} Kamata, Y., Koyama, K., Tsuboi, Y.,
       \& Yamauchi, S.\ 1997, \pasj, 49, 461
\bibitem[Kamazaki et al.(2001)]{Kamazaki2001} Kamazaki, T., Saito, M., Hirano, N.,
       \& Kawabe, R. 2001, \apj, 548, 278
\bibitem[Kamazaki et al.(2003)]{Kamazakietal2003} Kamazaki, T., Saito, M., Hirano, N., Umemoto, T.,
       \& Kawabe, R. 2003, \apj, 584, 357
\bibitem[Kamegai et al.(2003)]{Kamegai2003} Kamegai, K., Ikeda, M.,
         Maezawa, H., Ito, T., Iwata, M., et al. 2003, \apj, 589, 378
\bibitem[Kessler-Silacci et al.(2006)] {Kessler2006} Kessler-Silacci, J. E. et al. 2006, \apj, 639, 275
\bibitem[Khanzadyan et al.(2004)]{Khanzadyanetal2004} Khanzadyan, T., Gredel, R., Smith, M. D.,
       \& Stanke, T. 2004, \aap, 426, 171
\bibitem[Knude \& H{\o}g(1998)]{KnudeandHog1998} Knude, J. \& H{\o}g, E. 1998, \aap, 338, 897
\bibitem[Kuan et al.(2004)]{Kuan2004} Kuan, Y.-J., Huang, H.-C.,
             Charnley, S. B., Hirano, N.,
             Takakuwa, S., et al. 2004, \apj, 616, L27
\bibitem[Kulesa et al.(2005)] {Kulesa2005} Kulesa, C. A., Hungerford, A. L., Walker, C. K., Zhang, X.,
       \& Lane, A. P. 2005, \aj, 625, 194
\bibitem[Lada \& Wilking(1984)]{LadaandWilking1984} Lada, C. J. \& Wilking, B. A. 1984, \apj, 287, 610
\bibitem[Lahuis et al.(2006)] {Lahuis2006} Lahuis, F., van Dishoeck, E. F., Boogert, A. C. A.,  Pontoppidan, K. M.,
         Blake, G. A. et al. 2006, \apj, 636, L145
\bibitem[Lahuis et al.(2007)] {Lahuis2007} Lahuis, F., van Dishoeck, E. F., Blake, G. A., Evans, N. J.,
         Kessler-Silacci, J. E., \& Pontoppidan, K. M. 2007, \apj, 665, 492
\bibitem[Leous et al.(1991)]{Leousetal1991} Leous, J. A., Feigelson, E. D.,  Andr{\' e}, Ph.,
       \& Montmerle, T. 1991, \apj, 379, 683
\bibitem[Lis et al.(2002)]{Lis2002} Lis, D. C., Gerin, M., Phillips, T. G.,
       \& Motte, F. 2002, \apj, 569, 322
\bibitem[Liseau et al.(1999)]{Liseau1999} Liseau, R., White, G. J., Larsson, B.,
        Sidher, S., Olofsson, G., Kaas, A., Nordh, L., Caux, E., Lorenzetti, D., Molinari, S.,
        Nisini, B., \& Sibille, F. 1999, \aap, 344, 342
\bibitem[Loinard et al.(2007)]{Loinard2007} Loinard, L., Chandler, C., Rodr{\' i}guez, L., D'Alessio, P.,
        Brogan, C., Wilner, D., \& Ho, P. 2007, \apj, 670, 1353
\bibitem[Loinard et al.(2008)]{Loinard2008} Loinard, L., Torres, R. M., Mioduszewski, A. J.,
       \& Rodr{\' i}guez, L. F. 2008, \apj, 675, L29
\bibitem[Lombardi, Lada, \& Alves(2008)]{Lombardi2008} Lombardi, M., Lada, C. J., \& Alves, J. 2008, \aap, 480, 785
\bibitem[Looney, Mundy, \& Welch(2000)]{Looney2000} Looney, L. W., Mundy, L. G.,
       \& Welch, W. J. 2000, \apj, 529, 477
\bibitem[Loren(1989a)]{Loren1989a} Loren, R. B. 1989a, \apj, 338, 902
\bibitem[Loren(1989b)]{Loren1989b} Loren, R. B. 1989b, \apj, 338, 925
\bibitem[Loren \& Wootten(1986)]{LorenandWootten1986} Loren, R. B. \& Wootten, H. A.
        1986, \apj, 306, 142
\bibitem[Loren, Wootten, \& Wilking(1990)] {LorenWoottenandWilking1990}  Loren, R. B., Wootten, H. A.,
        \& Wilking, B. A. 1990, \apj, 365, 269
\bibitem[Luhman, Liebert, \& Rieke(1997)]{LuhmanLiebertandRieke1997} Luhman, K. L., Liebert, J.,
        \& Rieke, G. H. 1997, \apj, 489, L165
\bibitem[Luhman \& Rieke(1999)]{LuhmanandRieke1999} Luhman, K. L. \& Rieke, G. H.
        1999, \apj, 525, 440
\bibitem[Makarov(2007)] {Makarov2007} Makarov, V. V. 2007, \apj, 670, 1225
\bibitem[Mamajek(2008)]{Mamajek2007} Mamajek, E. E. 2008, Astron. Nachr., 329, 10
\bibitem[Mart{\' i}n et al.(1998)]{Martinetal1998} Mart{\' i}n, E. L., Montmerle, T.,
        Gregorio-Hetem, J., \& Casanova, S. 1998, \mnras, 300, 733
\bibitem[Miller \& Scalo(1979)]{MillerandScalo1979} Miller, G. E. \& Scalo, J. M. 1979, \apjs, 41, 513
\bibitem[Mizuno et al.(1989)]{Mizuno1989} Mizuno, A., Nozawa, S., \& Fukui, Y. 1989, unpublished data
\bibitem[Mizuno et al.(1990)]{Mizuno1990} Mizuno, A., Fukui, Y., Nozawa, S.,
        \& Takano, T. 1990, \apj, 356, 184
\bibitem[Montmerle et al.(1983)]{Montmerle1983} Montmerle, T., Koch-Miramond, L.,
        Falgarone, E., \& Grindlay, J. E. 1983, \apj, 269, 182
\bibitem[Motte, Andr{\' e}, \& Neri(1998)]{MotteAndreandNeri1998} Motte, F.,  Andr{\' e}, Ph.,
       \& Neri, R. 1998, \aap, 336, 150
\bibitem[Mundy et al.(1992)]{Mundyetal1992} Mundy, L. G., Wootten, A., Wilking, B. A., Blake, G. A.,
       \& Sargent, A. I. 1992, \apj, 385, 306
\bibitem[Natta et al.(2002)]{Nattaetal2002} Natta, A., Testi, L., Comer\'on, F., Oliva, E., D'Antona, F.,
        Baffa, C., Comoretto, G., \& Gennari, S. 2002, \aap, 393, 597
\bibitem[Natta et al.(2004)]{Nattaetal2004} Natta, A., Testi, L., Muzerolle, J., Randich, S., Comer\'on, F.,
       \& Persi, P. 2004, \aap, 424, 603
\bibitem[Natta, Testi, \& Randich(2006)]{Natta2006} Natta, A. Testi, L., \& Randich, S. 2006, \aap, 452, 245
\bibitem[Nozawa et al.(1991)]{Nozawaetal1991} Nozawa, S., Mizuno, A., Teshima, Y., Ogawa, H.,
       \& Fukui, Y. 1991, \apjs, 77, 647
\bibitem[Nutter, Ward-Thompson, \& Andr{\' e}(2006)]{Nutter2006} Nutter, D., Ward-Thompson, D., \& Andr{\' e}, Ph.
        2006, \mnras, 386, 1833
\bibitem[Ozawa, Grosso, \& Montmerle(2005)]{Ozawa2005} Ozawa, H., Grosso, N., \& Montmerle, T.
        2005, \aap, 429, 963
\bibitem[Padoan \& Nordlund(2002)]{Padoan2002} Padoan, P. \& Nordlund, A. 2002, \apj, 576, 870
\bibitem[Padgett et al.(2008)]{Padgett2008} Padgett, D., Rebull, L. M., Stapelfeldt, K. R., Chapman, N. L.,
        Lai, S.-P. et al. 2008, \apj, 672, 1013
\bibitem[Patience, Akeson, \& Jensen(2008)]{Patience2008} Patience, J., Akeson, R. L.,
        \& Jensen, E. L. N. 2008, \apj, 677, 616
\bibitem[Phelps \& Barsony(2004)]{PhelpsandBarsony2004} Phelps, R. \& Barsony, M. 2004, \aj, 127, 420
\bibitem[Phillips, Lonsdale, \& Feigelson(1991)]{Phillips1991} Phillips, R. B., Lonsdale, C. J.,
       \& Feigelson, E. D. 1991, \apj, 382, 261
\bibitem[Pontoppidan(2006)] {Pontoppidan2006} Pontoppidan, K. M. 2006, \aap, 453, L47
\bibitem[Pontoppidan et al.(2005)] {Pontoppidan2005} Pontoppidan, K. M., Dullemond, C. P., van Dishoeck, E. F.,
        Blake, G. A., Boogert, A. C. A.  et al. 2005, \apj, 622, 463
\bibitem[Pontoppidan et al.(2007)] {Pontoppidan2007} Pontoppidan, K. M., Stapelfeldt, K. R., Blake, G. A.,
       van Dishoeck, E. F., \& Dullemond, C. P. 2007, \apj, 658, L111
\bibitem[Preibisch et al.(2002)]{Preibischetal2002} Preibisch, T., Brown, A. G. A.,
        Bridges, T., Guenther, E., \& Zinnecker, H. 2002, \aj, 124, 404
\bibitem[Pudritz et al.(1996)]{Pudritzetal1996} Pudritz, R. E., Wilson, C. E., Carlstrom, J. E., Lay, O. P., Hills, R. E.,
       \& Ward-Thompson, D. 1996, \apj, 470, L123
\bibitem[Ratzka, K\"{o}hler, \& Leinert(2005)]{Ratzka2005} Ratzka, T., K\"{o}hler, R.,
       \& Leinert, Ch. 2005, \aap, 437, 611
\bibitem[Reipurth \& Zinnecker(1993)]{ReipurthandZinnecker1993} Reipurth, B. \& Zinnecker, H. 1993, \aap, 278, 81
\bibitem[Remijan \& Hollis(2006)]{Remijan2006} Remijan, A. J. \& Hollis, J. M. 2006, \apj, 640, 842
\bibitem[Ressler \& Barsony(2001)]{ResslerandBarsony2001} Ressler, M. E. \& Barsony, M. 2001, \aj, 121, 1098
\bibitem[Ressler \& Barsony(2003)]{ResslerandBarsony2003} Ressler, M. E. \& Barsony, M. 2003, \apj, 584, 832
\bibitem[Richichi et al.(1994)]{Richichietal1994} Richichi, A., Leinert, Ch., Jameson, R.,
       \& Zinnecker, H. 1994, \aap, 287, 145
\bibitem[Ridge et al.(2006)]{Ridge2006} Ridge, N. A., Di Francesco, J., Kirk, H., Li, D., Goodman, A. A.
         et al. 2006, \aj, 131, 2921
\bibitem[Rieke \& Rieke(1990)]{RiekeandRieke1990} Rieke, G. H. \& Rieke, M. J. 1990, \apj, 362, L21
\bibitem[Robitaille et al.(2006)]{Robitaille2006} Robitaille, T. P., Whitney, B. A., Indebetouw, R., Wood, K.,
       \& Denzmore, P. 2006, \apjs, 167, 256
\bibitem[Rodr{\' i}guez et al.(2005)]{Rodriguez2005} Rodr{\' i}guez, L. F., Loinard, L., D'Alessio, P., Wilner, D. J.,
       \& Ho, P. T. P. 2005, \apj, 621, L133
\bibitem[Rydgren(1980)]{Rydgren1980} Rydgren, A. E. 1980, \aj, 85, 438
\bibitem[Sch\"{o}ier et al. (2004)]{Schoier2004} Sch\"{o}ier, F. L., Jorgensen, J. K., van Dishoeck, E. F.,
       \& Blake, G. A. 2004, \aap, 418, 185
\bibitem[Sciortino et al.(2006)]{Sciortino2006} Sciortino, S., Pillitteri, I., Damiani, F.,
              Flaccomio, E., Micela, G., et al. 2006, in Proceedings of {\it The
        X-ray Universe 2005}, ed. A. Wilson, ESA SP-604, Volume 1, 111
\bibitem[Sekimoto et al.(1997)]{Sekimotoetal1997} Sekimoto, Y., Tatematsu, K., Umemoto, T.,
         Koyama, K., Tsuboi, Y., Hirano, N., \& Yamamoto, S. 1997, \apj, 489, L63
\bibitem[Simon et al.(1987)]{Simonetal1987} Simon, M., Howell, R. R., Longmore, A. J., Wilking, B. A., Peterson, D. M.,
       \& Chen, W.-P. 1987, \apj, 320, 344
\bibitem[Simon et al.(1995)]{Simonetal1995} Simon, M., Ghez, A. M., Leinert, Ch., Cassar, L., Chen, W. P.,
        Howell, R. R., Jameson, R. F., Matthews, K., Neugebauer, G., \& Richichi, A. 1995, \apj, 443, 625
\bibitem[Simpson et al.(2004)]{Simpsonetal2004} Simpson, C. M., Claussen, M. J., Wilking, B. A., Wootten, H. A.,
       \& Marvel, K. B. 2004, \baas, 36, 1567
\bibitem[Slesnick et al.(2006)]{Slesnick2006} Slesnick, C. L., Carpenter, J. M.,
        \& Hillenbrand, L. A. 2006, \aj, 131, 3016
\bibitem[Smith et al.(2005)]{Smithetal2005} Smith, M. D., Gredel, R.,  Khanzadyan, T.,
       \& Stanke, T. 2005, Mem. S. A. It., 76, 247
\bibitem[Stanke et al.(2006)]{Stankeetal2006} Stanke, T., Smith, M. D., Gredel, R.,
        \& Khanzadyan, T. 2006, \aap, 447, 609
\bibitem[Stark et al.(2004)]{Starketal2004} Stark, R., Sandell, G., Beck, S., Hogerheijde, M.,
         van Dishoeck, E., van der Wal, P., van der Tak, F., Sch\"{a}fer, F., Melnick, G., Ashby, M.
          \& de Lange, G. 2004, \apj, 608, 341
\bibitem[Strom, Kepner, \& Strom(1995)]{StromKepnerStrom1995} Strom, K. M., Kepner, J.,
       \& Strom, S. E. 1995, \apj, 438, 813
\bibitem[Struve \& Rudkj{\o}bing(1949)]{StruveandRudkjobing1949} Struve, O. \& Rudkj{\o}bing, M. 1949, \apj, 109, 92
\bibitem[Terebey et al.(2001)]{Terebey2001} Terebey, S., Van Buren, D., Hancock, T., Padgett, D.,
       \& Brundage, M. 2001, in ASP Conf. Ser. 243, {\it From Darkness to Light: Origin and Evolution of Young
        Stellar Clusters}, ed. T. Montmerle \& Ph. Andr{\' e}, 243
\bibitem[Testi et al.(2002)]{Testietal2002} Testi, L., Natta, A., Oliva, E., D'Antona, F.,  Comer\'on, F.,
        \& Baffa, C. 2002, \apj, 571, L155
\bibitem[Umemoto et al.(2002)]{Umemotoetal2002} Umemoto, T., Kamazaki, T., Sunada, K., Kitamura, Y.,
       \& Hasegawa, T. 2002, in {\it The Proceedings of the IAU 8th Asian-Pacific Regional Meeting},
        Volume II, ed. S. Ikeuchi, J. Hearnshaw, and T. Hanawa, 229
\bibitem[Vrba et al.(1975)]{Vrbaetal1975} Vrba, F. J., Strom, S. E., Strom, K. M.,
        \& Grasdalen, G. L. 1975, \apj, 197, 77
\bibitem[Vrba, Strom, \& Strom(1976)]{Vrbaetal1976} Vrba, F. J., Strom, S. E.,
        \& Strom, K. M. 1976, \aj, 81, 958
\bibitem[Vrba(1977)]{Vrba1977} Vrba, F. J. 1977, \aj, 82, 198
\bibitem[Walker et al.(1988)]{Walkeretal1988} Walker, C. K., Lada, C. J., Young, E. T.,
       \& Margulis, M. 1988, \apj, 332, 335
\bibitem[Walter et al.(1994)]{Walteretal1994} Walter, F. M., Vrba, F. J., Mathieu, R. D., Brown, A.,
        \& Myers, P. C. 1994, \aj, 107, 692
\bibitem[Ward-Thompson et al.(1989)]{Ward-Thompsonetal1989} Ward-Thompson, D., Robson, E. I.,
        Whittet, D. C. B., Gordon, M. A., Walther, D. M., \& Duncan, W. D. 1989, \mnras, 241, 119
\bibitem[Whittet(1974)]{Whittet1974} Whittet, D. C. B. 1974, \mnras, 168, 361
\bibitem[Wilking(1991)]{Wilking1991} Wilking, B. A. 1991,
        in {\it Star Formation in Southern Molecular Clouds}, ed. B. Reipurth, ESO Scientific Report No. 11, 159
\bibitem[Wilking et al.(2001)]{Wilkingetal2001} Wilking, B. A., Bontemps, S., Schuler, R. E., Greene, T. P.,
       \& Andr{\' e}, Ph. 2001, \apj, 551, 357
\bibitem[Wilking, Greene, \& Meyer(1999)]{WilkingGreeneandMeyer1999} Wilking, B. A., Greene, T. P.,
       \& Meyer, M. R. 1999, \aj, 117, 469
\bibitem[Wilking \& Lada(1983)]{WilkingandLada1983} Wilking, B. A. \& Lada, C. J. 1983, \apj, 274, 698
\bibitem[Wilking, Lada, \& Young(1989)]{WilkingLadaandYoung1989} Wilking, B. A., Lada, C. J.,
       \& Young, E. T. 1989, \apj, 340, 823
\bibitem[Wilking, Meyer, \& Greene(2008)]{Wilking2007} Wilking, B. A., Meyer, M. R.,
       \& Greene, T. P. 2008, in preparation
\bibitem[Wilking et al.(2005)]{Wilking2005}Wilking, B. A., Meyer, M. R., Robinson, J. G.,
       \& Greene, T. P. 2005, \aj, 130, 1733
\bibitem[Wilking, Schwartz, \& Blackwell(1987)]{WilkingSchwartzandBlackwell1987} Wilking, B. A., Schwartz, R. D.,
       \& Blackwell, J. H. 1987, \aj, 94, 106
\bibitem[Wilking et al.(1997)]{Wilkingetal1997} Wilking, B. A., Schwartz, R. D., Fanetti, T. M.,
       \& Friel, E. D. 1997, \pasp, 109, 549
\bibitem[Williams et al.(1995)]{Williamsetal1995} Williams, D. M., Comer\'on, F., Rieke, G. H.,
       \& Rieke, M. J. 1995, \apj, 454, 144
\bibitem[Wilson et al.(1999)]{Wilson1999} Wilson, C. D., Avery, L. W., Fich, M., Johnstone, D.,
        Joncas, G., Knee, L. B. G., Matthews, H. E., Mitchell, G. F., Moriarty-Schieven, G. H.,
       \& Pudritz, R. E. 1999, \apj, 513, 139
\bibitem[Wootten \& Loren(1987)]{WoottenandLoren1987} Wootten, H. A. \& Loren, R. B. 1987, \apj, 317, 220
\bibitem[Wootten(1993)]{Wootten1993}  Wootten, A. 1993, in {\it Astrophysical Masers}, ed. A. W. Clegg
       \& G. E. Nedoluha, (Berlin/Heidelberg: Springer), 315
\bibitem[Wootten et al.(1999)]{Woottenetal1999} Wootten, A., Claussen, M., Marvel, K.,
       \& Wilking, B. 1999, in {\it The Physics and Chemistry of the Interstellar Medium}, Proceedings of the
        3rd Cologne-Zermatt Symposium, eds. V. Ossenkopf, J. Stutzki, and G. Winnewisser, 295
\bibitem[Wu et al.(2002)]{Wu2002} Wu, J., Wang, M., Yang, J., Deng, L., \& Chen, J. 2002, \aj, 123, 1986
\bibitem[Ybarra et al.(2006)] {Ybarra06} Ybarra, J. E., Barsony, M., Haisch, K. E., Jarrett, T. H., Sahai, R.,
        \& Weinberger, A. J. 2006, \apj, 647, L159
\bibitem[Yeh et al.(2008)] {Yeh2007} Yeh, S., Hirano, N., Bourke, T., Ho, P., Lee, C.-F., Ohashi, N.,
        \& Takakuwa, S. 2008, \apj, 675, 454
\bibitem[Young, Lada, \& Wilking(1986)]{YoungLadaandWilking1986} Young, E. T., Lada, C. J.,
       \& Wilking, B. A., \apj, 304, L45
\bibitem[Young et al.(2006)]{Young2006} Young, K. E., Enoch, M. L.,
         Evans, N. J., II, Glenn, J., Sargent, A., et al. 2006, \apj, 644, 326
\bibitem[Zhang, Wootten, \& Ho(1997)]{Zhang1997} Zhang, Q., Wootten, A., \& Ho, P. T. P. 1997, \apj, 475, 713

\end{thebibliography}
\end{document}